\documentclass[iop]{emulateapj}
\usepackage{graphicx}
\usepackage{color}
\usepackage{amssymb}
\usepackage{amsmath}
\usepackage{epstopdf}
\usepackage{xspace}
\newcommand{\eref}{Equation~\ref}
\newcommand{\fref}{Figure~\ref}
\newcommand{\tref}{Table~\ref}
\newcommand{\sref}{Section~\ref}
\newcommand{\ssref}{Sections~\ref}

\newcommand{\nus}{{\em NuSTAR}\xspace}

\newcommand{\rxte}{{\em RXTE}\xspace}

\newcommand{\cts}{\ensuremath{{\rm cts\,s}^{-1}}\xspace}
\def\arcsec{\ensuremath{\,^{\prime\prime}}\xspace}
\def\arcmin{\ensuremath{\,^{\prime}}\xspace}

\newcommand{\inprep}{{\em in prep.}\xspace}
\newcommand{\submit}{{\em submitted}\xspace}

\newcommand{\deadt}{\ensuremath{\tau_d}\xspace}
\newcommand{\tlag}{\ensuremath{t_{\rm lag}}\xspace}
\newcommand{\rin}{\ensuremath{r_{{\rm in}}}\xspace}
\newcommand{\rdet}{\ensuremath{r_{{\rm det}}}\xspace}
\newcommand{\ft}{\ensuremath{\mathcal{F}}\xspace}
\newcommand{\rms}{\ensuremath{{\rm r.m.s.}}\xspace}

\newcommand{\cyg}{Cyg\,X-1\xspace}
\newcommand{\gx}{GX\,339$-$4\xspace}
\newcommand{\grs}{GRS\,1915$+$105\xspace}

\newcommand{\berk}{4}
\newcommand{\bamb}{5}
\newcommand{\dtu}{6}
\newcommand{\lawr}{7}
\newcommand{\camb}{8}
\newcommand{\durh}{9}
\newcommand{\colu}{10}
\newcommand{\mich}{11}
\newcommand{\texas}{12}
\newcommand{\gsfcb}{13}
\newcommand{\jpl}{14}
\newcommand{\uva}{15}
\newcommand{\gsfc}{16}

\shorttitle{Timing of BHs with \nus}
\shortauthors{Bachetti {\em et al.}}

\begin{document}
\title{No Time for Dead Time: Timing analysis of bright black hole binaries with \nus}

\author{Matteo Bachetti\altaffilmark{1,2}}\email{matteo.bachetti@irap.omp.eu}
\author{Fiona A. Harrison\altaffilmark{3}}
\author{Rick Cook\altaffilmark{3}}
\author{John Tomsick\altaffilmark{\berk}}
\author{Christian Schmid\altaffilmark{\bamb}}
\author{Brian W. Grefenstette\altaffilmark{3}}
\author{Didier Barret\altaffilmark{1,2}}
\author{Steven E. Boggs\altaffilmark{\berk}}
\author{Finn E. Christensen\altaffilmark{\dtu}}
\author{William W. Craig\altaffilmark{\berk,\lawr}}
\author{Andrew C. Fabian\altaffilmark{\camb}}
\author{Felix F\"urst\altaffilmark{3}}
\author{Poshak Gandhi\altaffilmark{\durh}}
\author{Charles J. Hailey\altaffilmark{\colu}}
\author{Erin Kara\altaffilmark{\camb}}
\author{Thomas J. Maccarone\altaffilmark{\texas}}
\author{Jon M. Miller\altaffilmark{\mich}}
\author{Katja Pottschmidt\altaffilmark{\gsfcb}}
\author{Daniel Stern\altaffilmark{\jpl}}
\author{Phil Uttley\altaffilmark{\uva}}
\author{Dominic J. Walton\altaffilmark{3}}
\author{J\"orn Wilms\altaffilmark{\bamb}}
\author{William W. Zhang\altaffilmark{\gsfc}}

\altaffiltext{1}{Universit\'e de Toulouse; UPS-OMP; IRAP; Toulouse, France}
\altaffiltext{2}{CNRS; Institut de Recherche en Astrophysique et Plan\'etologie; 9 Av. colonel Roche, BP 44346, F-31028 Toulouse cedex 4, France}
\altaffiltext{3}{Cahill Center for Astronomy and Astrophysics, Caltech, Pasadena, CA 91125}
\altaffiltext{\berk}{Space Sciences Laboratory, University of California, Berkeley, CA 94720, USA}
\altaffiltext{\bamb}{Dr. Karl-Remeis-Sternwarte and ECAP, Sternwartstr. 7, 96 049 Bamberg, Germany}
\altaffiltext{\dtu}{DTU Space, National Space Institute, Technical University of Denmark, Elektrovej 327, DK-2800 Lyngby, Denmark}
\altaffiltext{\lawr}{Lawrence Livermore National Laboratory, Livermore, CA 94550, USA}
\altaffiltext{\camb}{Institute of Astronomy, University of Cambridge, Madingley Road, Cambridge CB3 0HA, UK}
\altaffiltext{\durh}{Department of Physics, Durham University, South Road DH1 3LE, UK}
\altaffiltext{\colu}{Columbia Astrophysics Laboratory, Columbia University, New York, NY 10027, USA}
\altaffiltext{\mich}{Department of Astronomy, University of Michigan, 500 Church Street, Ann Arbor, MI 48109-1042, USA}
\altaffiltext{\texas}{Department of Physics, Texas Tech University, Lubbock, TX 79409, USA}
\altaffiltext{\gsfcb}{CRESST, UMBC, and NASA GSFC, Code 661, Greenbelt, MD 20771, USA}
\altaffiltext{\jpl}{Jet Propulsion Laboratory, California Institute of Technology, Pasadena, CA 91109, USA}
\altaffiltext{\uva}{Anton Pannekoek Institute, University of Amsterdam, Science Park 904, 1098 XH Amsterdam, The Netherlands}
\altaffiltext{\gsfc}{NASA Goddard Space Flight Center, Greenbelt, MD 20771, USA}

\begin{abstract}
Timing of high-count rate sources with the \nus Small Explorer Mission requires specialized analysis techniques.  \nus was primarily designed for
spectroscopic observations of sources with relatively low count-rates rather than for timing analysis of bright objects.  The instrumental dead time per event 
is relatively long ($\sim$2.5~msec), and varies by a few percent event-to-event.  
The most obvious effect is a distortion of the white noise level in the power density spectrum (PDS) that 
cannot be modeled easily with the standard techniques due to the variable nature of the dead time.
In this paper, we show that it is possible to exploit the presence of two completely independent focal planes and use the cross 
power density spectrum to obtain a good proxy of the white noise-subtracted PDS.   
Thereafter, one can use a Monte Carlo approach to estimate the remaining effects of dead time, 
namely a frequency-dependent modulation of the variance and a frequency-independent drop of the sensitivity to variability.
In this way, most of the standard timing analysis can be performed, albeit with a sacrifice in signal to noise relative to what would be
achieved using more standard techniques.
We apply this technique to \nus observations of the black hole binaries \gx, \cyg and \grs.

\end{abstract}

\keywords{X-rays: stars --- accretion, accretion disks --- black hole physics}

\maketitle

%
%

\section{Introduction}

Timing analysis, other than being an important diagnostic tool by itself \citep[see][for a review]{Vaughan13}, is particularly powerful for dissecting the inner accretion regions in black hole binaries and AGN when combined with spectral modeling.
Good examples of this combination in binaries are studies of the variability of a specific spectral component \citep[e.g. the iron K$_{\alpha}$ line and the reflection component, ][]{Revnivtsev+99,Gilfanov+00}, the study of the time lags between different energy bands and the comparison with the expected behavior of a given spectral component \citep[e.g.][]{Nowak+96,Papadakis+01,KoerdingFalcke04,Fabian+09,Gandhi+10,Uttley+11,Artigue+13}, the study of the covariance of the signal at multiple energies and the variability spectrum \citep[e.g.][]{Uttley+11,Jin+13}.
A review of reverberation lag measurements in binaries and AGN can be found in \citet{Uttley+14}.

The {\em Nuclear Spectroscopic Telescope ARray} (\nus; \citealt{nustar13}) mission has deployed the first hard X-ray (above 10\,keV) focusing satellite. 
It features two telescopes, focusing X-rays between 3 and 79\,keV onto two identical focal planes (usually called focal plane modules A and B, or FPMA and FPMB).
It has a field of view (FOV) of 12\arcmin$\times$12\arcmin and an angular resolution of 18\arcsec FWHM; (58\arcsec HPD).
These features, together with good spectral resolution in the iron K band ($\sim$0.4\,keV@6\,keV), have motivated a large number of observations of X-ray binaries. 
A particularly interesting result was the accurate modeling of the reflection component in several black hole (BH) binaries \citep{Miller+131915,Tomsick+14} resulting in 
improved constraints on black hole spin in several systems.
Such measurements have also been made by \nus for active galactic nuclei \citep[AGN; e.g.][]{Risaliti+13,Marinucci+14,Walton+141365}.  \nus , {\em XMM} and {\em Suzaku} have recently
measured reverberation lags in the AGN MCG-5-23-16 in both the iron line and Compton reflection hump together for the first time \citep{Zoghbi+14}.

From the point of view of timing, \nus has an advantage over other imaging satellites: 
the time resolution is 10\,$\mu s$, and so one can in principle study variability over the whole range of interesting frequencies in accreting systems without switching to an observing mode with decreased spectral resolution. 
The satellite has in fact been used to perform timing analysis on selected targets with very good results:
for example, it was used to study a number of accreting and rotation-powered pulsars,
permitting the detection of variable cyclotron resonant scattering features \citep{Fuerst+13,Fuerst+14vela}
and the discovery of strong hard time lags in the emission of another accreting pulsar (GS 0834-430; \citealt{Miyasaka+13}).
\nus was also used to detect the pulsations from a magnetar near the Galactic Center \citep{Mori+13} and to carry out a detailed timing study of this serendipitous source \citep{Kaspi+14}.
These pulsars are mostly slow rotators, with pulse periods above 1\,s.
Aperiodic timing has been measured in faint sources \citep[e.g.][]{Bachetti+13}, or in bright sources but only over the low-frequency part of the power spectrum \citep[e.g.][]{Natalucci+14}.

From the point of view of fast ($\nu>>1$\,Hz) timing studies of bright sources, however, \nus's default observing mode requires care to analyze properly.
Each X-ray incident on a focal plane produces a trigger, and the energy signal is read out immediately.   The read time is dependent on the number of triggered pixels, and so the dead time for each event is slightly variable (at the few percent level) with an average value of $\sim2.5$\,ms.
This dead time, together with  additional dead time produced on a separate cadence by housekeeping operations and vetoed events, can be accurately measured and its value over a second is stored in the housekeeping files of the observations.   In addition, the live time since the previous X-ray event is recorded in a column in the event lists (see \sref{sec:deadtime} for details).   For bright sources where the fraction of events vetoed by the anti coincidence system is small,  light curves can be accurately produced with any time binning (see Madsen et al 2014, \submit, for application of this technique to the Crab pulsar).   However, for timing analysis (e.g. the production of power density spectra), the dead time introduces spurious correlations between event arrival times that are reflected in a distortion of the power spectrum, making it difficult to subtract the Poisson noise level above $\sim50$\,Hz.
This effect is almost completely negligible for count rates up to $\sim1$\cts, but it is very pronounced in observations of Galactic X-ray binaries during outbursts, with typical incident count rates $\gtrsim 100$\,\cts over the full band.

In this Paper, we present a technique to avoid spurious noise introduced by dead time, in particular the distortion of the white noise level in the power spectrum.  We show how a slight modification to the methods commonly used for quasi-periodic oscillation (QPO) detection and power spectral fitting enables good-quality results at all frequencies.
We use these techniques to perform a basic timing analysis of several \nus observations of BH binaries (BHBs), whose spectral analysis is presented in other papers (\grs, \citealt{Miller+131915}; \cyg, \citealt{Tomsick+14}; \gx, F\"urst et al. \inprep).  We chose these BHBs as ideal objects with which to demonstrate the timing techniques, since their timing behavior has been extensively studied by 
other missions and instruments (see \sref{sec:analysis} for details). 

The paper is organized as follows:  in \sref{sec:deadtime} we discuss details \nus's dead time, and how it is measured on-board the instrument; in \ssref{sec:cpds} and~\ref{sec:simu} we describe how to use the Cross Power Density Spectrum to obtain a proxy of the PDS; in \ssref{sec:red} and~\ref{sec:analysis} we use these techniques to analyze the data from 
two black hole X-ray Binaries, Cyg X-1 and GX~339-4; and in \sref{sec:concl} we provide conclusions.   In the Appendix we present results from a timing analysis of GRS~1915+105, which was observed
during the on-orbit commissioning period. This observation suffered from spurious features associated with the cadence of instrument housekeeping functions and so requires a unique analysis not relevant to \nus\ science phase observations.

\section{Dead time in \nus.}\label{sec:deadtime}

\begin{figure*}[t]
\includegraphics[width=\linewidth]{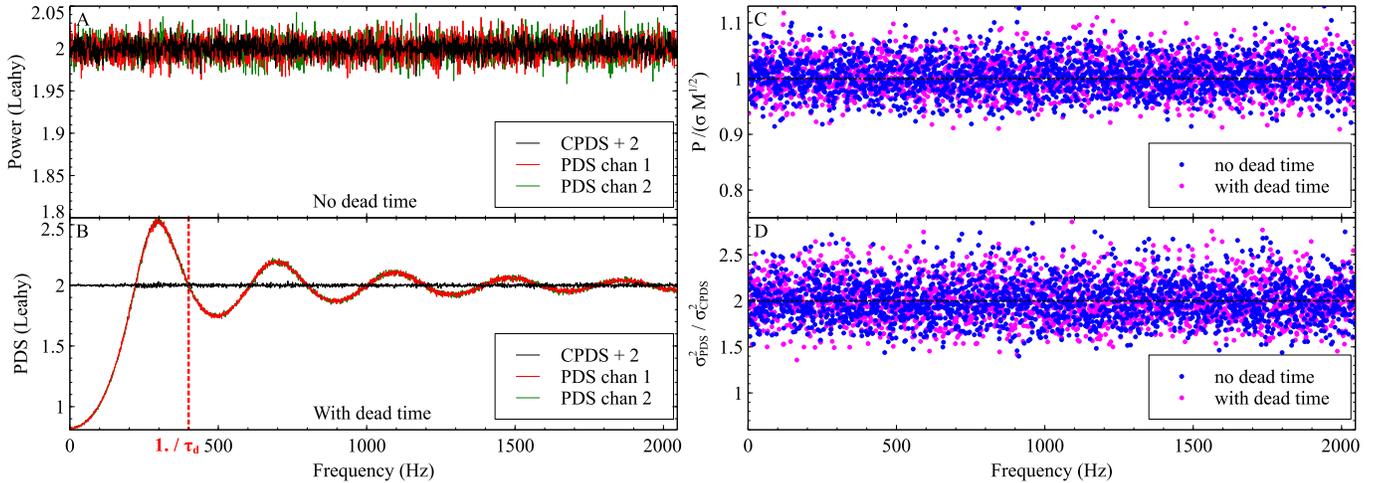}
\caption{(Left) The cospectrum and the PDS are compared in the case of pure Poisson noise, without (A) and with (B) dead time. The simulated incident count rate was 225 \cts.
The cospectrum mean is always 0. In these plots, it has been increased by 2 for display purposes.
The frequency 1/\deadt is indicated.
(Right) The usual relation between the PDS and its standard deviation ($\sigma = P /\sqrt{M}$, where $M$ is the number of averaged PDSs) holds with and without dead time (C).
Also, the variance of the cospectrum is half the variance of the PDS, in both cases (D).
}
\label{fig:cpdsvspds}
\end{figure*}

Dead time in the \nus\ instrument is produced when the focal plane module electronics are busy processing an event, when the shield veto prevents the focal plane from triggering (or interrupts an event that is being processed), and for some regular instrument housekeeping functions (the housekeeping produces less than 1 ms of dead time per second).   \nus\ has two independent telescopes, and the associated focal plane modules have independent processors, so that the dead time in the two modules is uncorrelated.
All of the events processed by the instrument are found in the unfiltered event file.
The cleaned event file is a subset of the events in the unfiltered event file where GTIs and additional screening have been applied to remove non-science quality events\footnote{See the NuSTARDAS user's guide available from the HEASARC (\href{http://heasarc.gsfc.nasa.gov/docs/nustar/analysis/}{http://heasarc.gsfc.nasa.gov/docs/nustar/analysis/}) for more information.}. 

For the production of light curves and flux measurements the \nus\ instrument has two operating modes, one for faint ($<$50 mCrab) sources, 
and one that allows highly accurate flux measurements even at very high
count rates and on arbitrarily short timescales. 
The fraction per second of dead time from events that are not processed, like shield vetoes, as well as that due to housekeeping functions
is saved in the housekeeping files on a one-second cadence.
For light curves binned in 1~s intervals (or a integer multiple thereof) multiplying by the live time stored in the housekeeping files
correctly accounts for dead time. 
In addition, the live time since the previous X-ray event is stored in the event lists in the ``PRIOR" column.
In the absence of events vetoed by the active anti-coincidence shield, this column would accurately reflect the true live time.     
In the operating mode used for most observations taken since in-orbit checkout, the events vetoed by the 
anti-coincidence shield are not telemetered to the ground to minimize data volume.    
For this mode adding the PRIOR column does not yield the proper
live time since vetoed events are not included.
However, for source count rates significantly greater than the veto rate the error is negligible (for the Crab it is only $\sim0.3$\%; see Madsen et al. 2014, \submit).
For sources of intermediate brightness, where the veto rate cannot be ignored, a fully tested and calibrated instrument mode exists that includes vetoed events in the data stream so that
fluxes and light curves can be produced with arbitrary accuracy.

For timing analyses that involve power spectral analysis, dead time produces systematic effects even if it is perfectly measured.
Dead time is classified as either paralyzable or non-paralyzable depending on
whether a photon hitting the detector during dead time produces new dead time or not.
The \nus\ readout architecture produces non-paralyzable dead time.
Its main effect is on recorded count rates:
if we have an incident rate of photons $\rin$,
the detected count rate will be approximately
\begin{equation}\rdet=\rin / (1 + \deadt \rin)
\end{equation}
 \citep[see, e.g.,][and references therein]{VDK89},
where \deadt is the dead time produced by each event, assuming that it is constant.
But dead time also alters the sensitivity to variable signals, acting as a frequency filter.
Power density spectra (PDS), in particular, are deformed to a ``wavy'' shape that depends on the magnitude of dead time and on count rates (see \fref{fig:cpdsvspds}).
Power at frequencies slightly above $1/\deadt$ is {\em quenched}, as there is a lack of events whose separation is less than $\deadt$, while there is a relatively higher rate above $\deadt$, and therefore the power at frequencies just below $1/\deadt$ is slightly {\em amplified}.
These ``waves'' have nodes at $1/\deadt$ and multiples thereof, where the power (in \citet{Leahy+83} normalization) is equal to 2, the value that it would have without dead time, and maxima and minima in between are given by the relative contribution of the quenching and amplification.
For frequencies $\nu << 1/\deadt$, the main effect is a general deficiency of events, and the power has a decreasing level that approaches $\approx 2(1 - \rin\deadt)^2$ \citep{Weisskopf85}.
Assuming that \deadt is the same for each event and that only source events
contribute to either non-paralyzable or paralyzable dead time, 
this distortion can be modeled precisely \citep[see][for an exhaustive treatment]{Vikhlinin+94,Zhang+95}.
Also, some statistical properties of the PDS hold in dead time-affected data.
For example, the standard deviation associated with the bin $i$ of the PDS is always equal to $P_i/\sqrt{M}$, where $P_i$ is the power in the bin $i$ and $M$ is the number of averaged PDSs (see \fref{fig:cpdsvspds}).

In \nus \deadt is not strictly fixed at the same value for all events, but varies by a few percent depending on the number of pixels that are triggered.
For this reason, the models available in literature do not correctly describe the dead time effects for this satellite:
the ``wavy'' behavior of the PDS shifts slightly, and to fully account for this effect and produce a white-noise subtracted PDS, a very precise modeling of the dead time would be required.
Since at high count rates the ``waves'' can be very prominent, any real variability feature such a QPO can easily be ``hidden'' and difficult to detect. 

As an additional complication, the models described above assume that dead time is produced completely by the recorded signal.
In \nus, additional dead time comes from events outside the source extraction region, from vetoed events, and from all events discarded for other reasons during the cleaning process in the pipeline (the step from unfiltered to cleaned event files).   In the following, we neglect the effect of vetoed events, since their dead time ($\sim20\,\mu$s) has characteristic frequency $1/\deadt\sim50$\,kHz, much higher than science events, and their total contribution to dead time is small.
We instead present a method that permits construction of a proxy of a white-noise subtracted PDS, regardless of the count rate and the ratio between source and background (or spurious) events. 

\section{The Cross Power Density Spectrum as an improved Power Spectrum}\label{sec:cpds}

\nus has two completely independent focal plane modules (each containing four detectors) that are read out by separate microprocessors.  It is therefore possible in principle to obtain the same information given by a PDS through the so-called Cross Power Density Spectrum (CPDS; for more details see \citealt{BendatPiersol11}): instead of considering the PDSs in the two individual focal planes:
\begin{equation}
P_{i}(\nu) = \ft_{i}^*(\nu)\ft_{i}(\nu) \; (i = A, B)
\end{equation}
where $\ft_{i}$ indicates the Fourier transform of the lightcurve detected by the focal plane $i$ and $\nu$ is the frequency,
one multiplies the complex conjugate of one Fourier transform with the other Fourier transform:
\begin{equation}
C(\nu) = \ft_{A}^*(\nu)\ft_{B}(\nu).
\end{equation}
The CPDS is often used in other contexts to obtain information on the correlation between the signal in two energy bands.
It is a complex quantity:
its real part is also called the {\em cospectrum} and gives a measure of the signal that is {\em in phase} between the two channels;
its imaginary part, or {\em quadrature spectrum}, gives instead a measure of the {\em off-phase} signal.
Therefore, in principle, it should be possible to eliminate all variability that is not related between the two light curves, including the effects of dead time, by just considering the cospectrum (the real part of the CPDS).
In \fref{fig:cpdsvspds} we show the statistical properties of the cospectrum in the case of pure Poisson noise.  In both the dead time affected and in the zero-dead time cases the cospectrum mean value is 0 (in \fref{fig:cpdsvspds}, it has been shifted to 2 for graphical reasons).
This is a big advantage, as this is independent of whether the dead time is constant or not (since the distribution of dead time is also independent between the two detectors), and therefore it is not necessary to conduct complicated studies of the dead time distribution in order to obtain a white noise-subtracted cospectrum, as opposed to a PDS where this procedure would be needed.

Another important point is that the standard deviation $\sigma_{\rm CPDS}$ of the cospectrum is linked to the standard deviations ($\sigma_{\rm PDS}$) of the PDSs of the two light curves. 
In Figure~1, we show that the ratio between $\sigma_{\rm CPDS}$ and the geometric average of the standard deviations of the two PDSs is close to $\sqrt{2}$ regardless of the frequency. 
This is very convenient, as it allows us to assign to the cospectrum bin $c_i$ an uncertainty of $\bar{P_i}/\sqrt{2M}$, where $\bar{P_i}$ is the geometric average of the two PDSs and $M$ is the number of averaged PDSs.
The argument (or angle) of the CPDS is called {\em phase lag};
if we divide it by $2\pi\nu$ we obtain the {\em time lag} \tlag, that is a measure of the time shift between the two channels.
In our case, since we are using two lightcurves in the same energy bands, we expect any time lag to be zero.
But time lags between different energy bands have been used as indicators of how a signal produced in an emitting region is reprocessed in other regions:
for example, if the disk emission is Comptonized in a corona,
or when the signal from a region is reflected or propagated in another region,
this time delay between the different emissions can be detected through time lags (see, e.g., \citealt{BendatPiersol11,Nowak+99, Uttley+14} and references therein).
We briefly discuss time lags between different energy bands in the following sections,
but for the most part our analysis will be performed on two channels in the same energy band, and the only important quantity for our treatment will be the cospectrum.

\begin{figure}[t]
\includegraphics[width=\columnwidth]{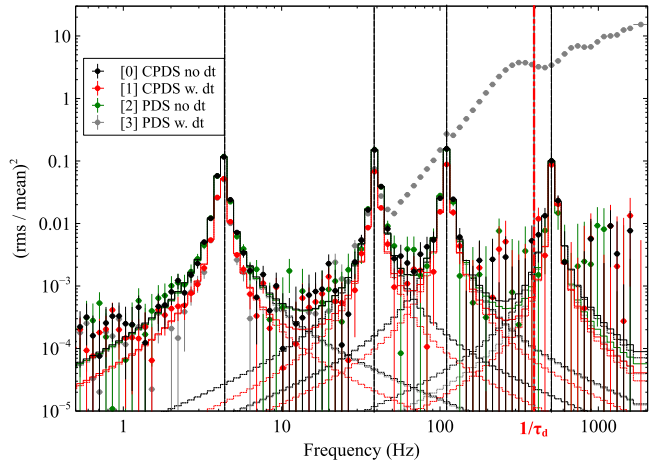}
\caption{QPOs of equal Q factor (20) and \rms amplitude (8\%) at different frequencies. 
The units in this plot, and in all following power spectra or cospectra if not stated otherwise, are Power$\times$Frequency.
Grey points show the standard, dead time-affected PDS.
The white noise level to subtract in the dead time-affected datasets was calculated between 10 and 20\,Hz (minimum between two QPOs),
while in the dead time-free case we subtracted the theoretical level (2 in Leahy normalization).}
\label{fig:qpos}
\end{figure}


\section{Simulations}\label{sec:simu}

In order to provide a fully consistent treatment of dead time effects in the PDS and in the cospectrum,
we ran a large number of simulations. 
In each simulation we produced two event series containing variability, one for each FPM,
and analyzed the data with cospectrum and PDS before and after applying a dead time filter.
By doing so, we studied in detail the properties of the cospectrum and compared them to those of the PDS.
In the following paragraphs we will explain the procedure in more detail, and demonstrate that the cospectrum can be considered a very good proxy of a white noise-subtracted PDS, albeit with some corrections to account for in the measured \rms.

\subsection{Procedure}\label{sec:simutec}

\paragraph{Light curve generation} 
We used the procedure by \citet{DaviesHarte87}, introduced to astronomers by \citet{TimmerKoenig95}, to simulate light curves between two times $t_0$ and $t_1$,
from a number of model PDS shapes containing QPOs.
The sampling frequency of the light curves was at least four times higher than the maximum frequency of the variability components included in the simulation.
We normalized the light curves in order to have the desired mean count rate and total \rms variability (7-10\%).
In order to be able to later calculate PDSs with a given maximum timescale $T$ (see ``Calculation of the PDSs and the cospectrum'' below), we simulated light curves at least 10 times longer, following the prescriptions in \citet{TimmerKoenig95} to avoid aliasing.

\paragraph{Event list generation}
From every light curve, we generated two event lists, corresponding to the signal from the two focal planes.
Each event list was produced as follows:
first of all, we calculated the number $N_{save}$ of event times to be generated as a random sample from a Poisson distribution
centered on the number of total photons expected (summing up all expected lightcurve counts);
then, we generated $N_{save}$ events by using a {\em Monte Carlo acceptance-rejection method}.
This is a classical Monte Carlo technique; a more general treatment can be found in most textbooks on Monte Carlo methods \citep[e.g.][]{Gentle03}.
In our case, we used the following procedure:
(1) for every event we simulated an event time $t_e$ uniformly distributed between $t_0$ and $t_1$,
and an associated random amplitude (``probability'') value $A_e$ between 0 and the maximum of the light curve;
(2) we rejected all $t_e$ whose associated $A_e$ was higher than the light curve at $t_e$;
to avoid possible spurious effects given by a stepwise model light curve, we used a {\em cubic spline} interpolation to approximate the light curve between bins;
(3) we sorted the event list by $t_e$.
To simulate the effects of background (spurious events filtered out by the pipeline, events recorded outside the source regions, etc.), we also produced two background series, at constant average flux, one for each simulated source light curve.

\paragraph{Dead time filtering}
For each event list, we created a corresponding dead time-affected event list by applying a simple dead time filter:
for each event, we eliminated all events in the 2.5\,ms after it.
Source and background events contributed equally to dead time. 
We used different versions of the dead time filter by varying slightly the dead time
between events (of $\sim0.1$ms).
But the pernicious effect of variable dead time is mainly on white noise subtraction.
In our case (\fref{fig:cpdsvspds}), the cospectrum allows us to overcome this problem as its white noise is 0, and we verified that the other effects are not significantly different in the constant and variable dead time cases. 
In the following, we will treat the case with constant dead time.

\paragraph{Calculation of the PDSs and the cospectrum}
We divided each pair of event lists in segments of length $T$, calculated the PDS in each of segments and the CPDS from each pair of them.
We then averaged the PDSs and CPDSs from all segments.
The CPDS white noise level is already 0.
For the PDSs, used only in the ideal zero-dead time case, we subtracted the theoretical Poisson level (2 in Leahy normalization).
We then rebinned the PDSs and CPDSs, either by a fixed rebin factor or by averaging a larger number of bins at high frequencies, following approximately a geometric progression.
We finally multiplied the Leahy-normalized PDSs by $(B+S)/S^2$ (where B is the background count rate and S the mean source count rate) in order to obtain the squared \rms normalization often used in literature \citep[][]{BelloniHasinger90,Miyamoto+91}. 
The CPDS was instead multiplied by the factor $(\overline{B+S})/\bar{S}^2$,
where bars indicate the {\em geometric} averages of the count rates in each of the two event lists used to calculate it%
\footnote{We chose this normalization by analogy: if the PDS is multiplied by $(B+S)/S^2$, it's as if each Fourier transform contributing to it were multiplied by $((B+S)/S^2)^{1/2}$.
Since the contribution to the CPDS is by two separate Fourier transforms, we multiplied by $((B_A+S_A)/S_A^2)^{1/2}((B_B+S_B)/S_B^2)^{1/2}$, that is equivalent to the geometric averages described.}.

Finally, we calculated the cospectrum by taking the real part of the CPDS. 
As described above, we assigned to each final bin of the cospectrum $c_i$ an uncertainty calculated from the geometrical average of the PDSs in the two channels, divided by $\sqrt{2MW}$, where $M$ is the number of averaged spectra and $W$ is the number of subsequent bins averaged to obtain $c_i$.

\paragraph{Fitting procedure}
Cospectra do not need Poisson noise subtraction; for PDSs, we fitted a constant to the interval outside of the frequency range containing QPOs.
Since in the following paragraphs, in all our examples, we will be showing power spectra obtained by averaging more than 50 PDSs, we are in the Gaussian regime and a fit with  standard $\chi^2$ minimization routines is appropriate to the precision we are interested in \citep{VDK89,Barret+12}.

Then, we fitted the QPOs with a Lorentzian profile in XSPEC\footnote{We used both the standard interface of the program, and its Python bindings, PyXSPEC.} \citep{Arnaud96}.
Errors were calculated through a Monte Carlo Markov Chain, as those intervals where the $\chi^2$ of the fit with no frozen parameters increased by 1. 
According to \citet{Lewin+88}, the significance of the detection of QPOs is expected to be
\begin{equation}\label{eq:sig}
n_{\sigma}\simeq \frac{1}{2} \frac{S^2}{B+S} r^2 {\left(\frac{T}{\Delta\nu}\right)}^{1/2} 
\end{equation}
where $r$ is the \rms and $\Delta\nu$ is the equivalent width of the feature (for a Lorentzian, $\Delta\nu=\pi/2\times{\rm FWHM}$).
The exact proportionality factor depends on the definition of significance. 
In our case, the significance of QPOs was defined as the ratio between the amplitude of the Lorentzian and its error.
The significance calculated in this way gives a value $\sim2$ times lower than obtained by using the excess power \`a la \citet{Lewin+88} \citep[a factor 2 is expected due to the fact that the Gaussian errors are calculated over $\mathbf{R}$, and the excess power over $\mathbf{R}^+$; see][]{BoutelierThesis,Boutelier+09}, but the trend in \eref{eq:sig} holds provided that the variability is dominated by Poisson noise.

\subsection{Simulation results}

\begin{figure}[t]
\centering
\includegraphics[width=\columnwidth]{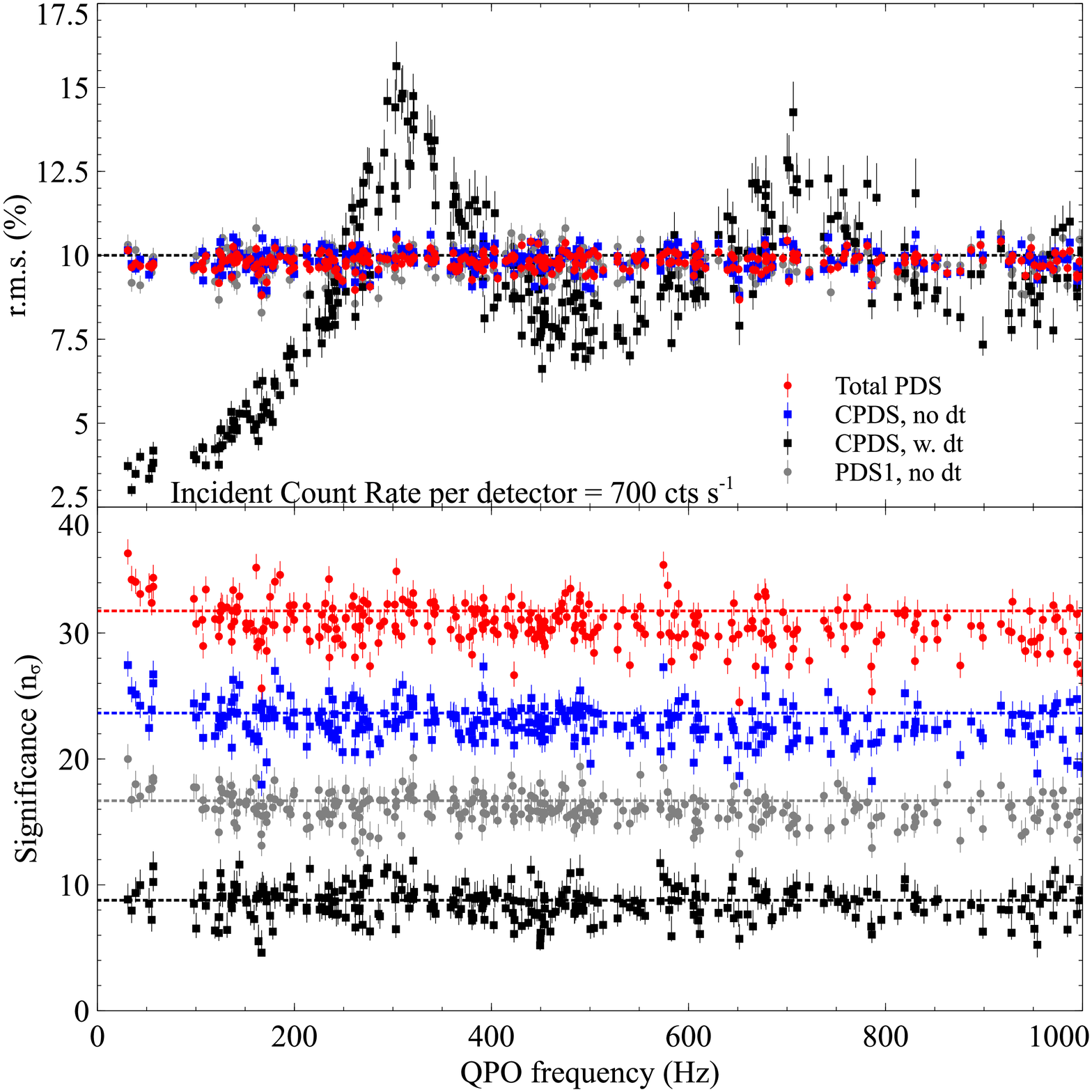}
\caption{(Top) Variation of the \rms of a QPO at different peak frequencies, measured with the various techniques and with and without dead time.
Each point represents a simulated QPO with \rms=10\% and FWHM=2\,Hz.  A total of 281 simulations were used for this plot.
(Bottom) Significance measured with each method.
The total PDS has about twice the significance of the single-module PDS in the no-dead time case, as expected, due to double number of photons.
The CPDS in the no-dead time case is a factor $\sim\sqrt{2}$ higher than the single PDS, and lower by the same amount than the total PDS.
The dead time-affected CPDS, instead, has a much lower level due to the lack of photons.
The decrease of significance does not depend on the frequency of the QPO, but only on count rate (see \fref{fig:cts}).
}
\label{fig:freq}
\end{figure}

\begin{figure}[t]
\centering
\includegraphics[width=\columnwidth]{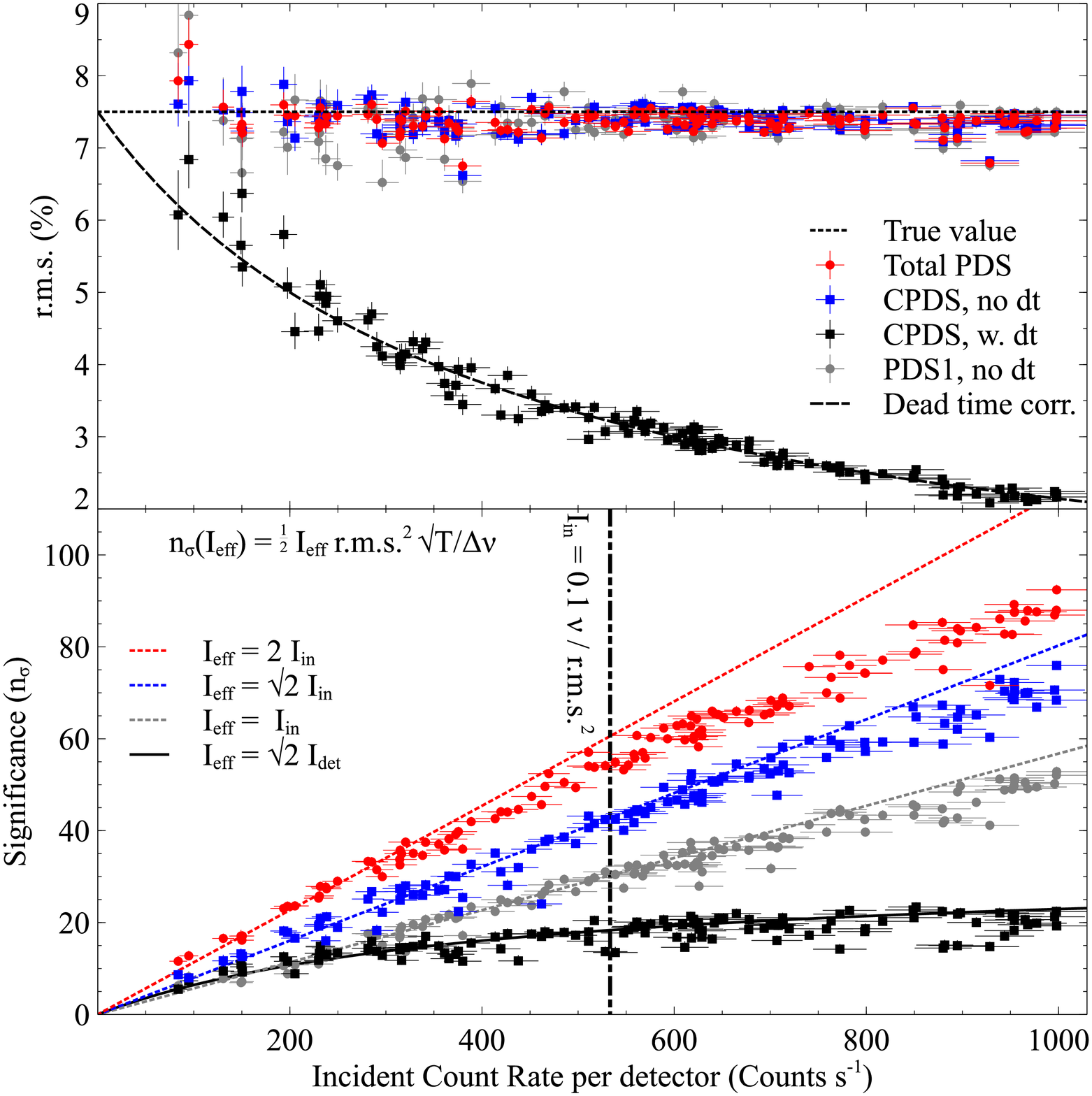}
\caption{Similar to \fref{fig:freq}, but with the centroid frequency of the QPO fixed at 30\,Hz and letting the count rate vary between 10 and 1000\,\cts.
A total of 118 simulations were used in this plot.
The line shows \eref{eq:drms}. It is not a fit, and describes remarkably well the data. 
In the bottom panel, we plot the detection significance for all the cases. 
The lines, again, are not fitted, they just show the theoretical prediction from \eref{eq:sig}.
All dead time-free cases are in good agreement with a linear increase with count rate below $\sim500\,\cts$, above which some curvature appears due to the departure from the quasi-Poissonian regime.
The dead time-affected case is in good agreement with \eref{eq:sig} if, instead of the incident count rate (dashed line), one uses the {\em observed} count rate (solid line; see \eref{eq:drms}).}
\label{fig:cts}
\end{figure}

\begin{figure}[t]
\centering
\includegraphics[width=\columnwidth]{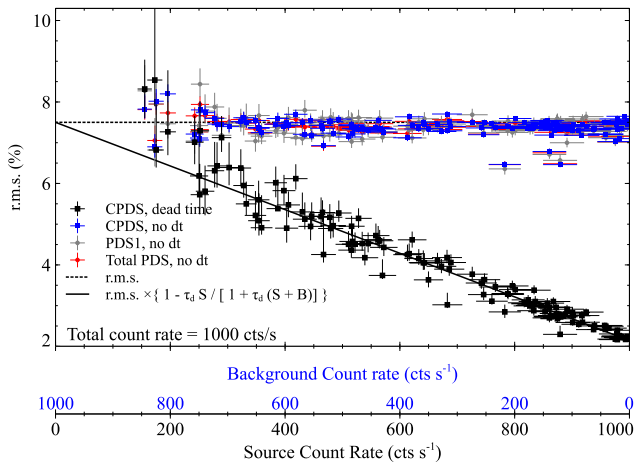}
\caption{The dependence of the \rms on the relative contribution of the source in a given energy range to the total count rate.
The 132 simulations that compose this plot show how the \rms drop is stronger if the source signal dominates the total signal, since the source signal contributes more to the total dead time.
The solid black line shows \eref{eq:drmstot}. It is not a fit.}
\label{fig:ctsback}
\end{figure}

\paragraph{First look} 
The simulation in \fref{fig:cpdsvspds} shows the comparison between the PDS and the cospectrum with and without dead time for pure Poisson noise.
From these simulations it is clear that the white noise-subtracted PDS and the cospectrum are equivalent in the case with no dead time.
It is immediately evident that the most problematic effect of dead time, the modulation of the white noise level, disappears in the cospectrum.
In the following paragraphs, we investigate the frequency and count rate dependence of these quantities in more detail.

\paragraph{Frequency dependence}
The general statistical properties of the PDS and the cospectrum are also very similar, both in the dead time-affected and in the zero-dead time case.
\fref{fig:cpdsvspds}, panel D, shows that the variance of the cospectrum and the PDS maintains a constant ratio
equal to 2 in both the clean and the dead time-affected datasets.
This makes it easy to calculate the variance of the cospectrum values for the subsequent analysis, by simply using the known properties of the PDS where the variance is just equal to the square of the power (in Leahy normalization).

Contrary to what might be imagined, it is possible to detect variability even at frequencies that are affected the most by dead time, i.e. those above $1/\deadt$. 
\fref{fig:qpos} shows that QPOs at all frequencies are detectable, albeit with some modulation of the observed \rms.
To measure this change of \rms, we simulated $\sim$500 lightcurves using the method above, each containing a single QPO with frequencies equally distributed between 5 and 1000\,Hz, \rms=10\% and FWHM=2\,Hz.
As explained above, from every light curve we obtained two event lists in order to simulate the signals from the two detectors.
We produced for every pair of event lists the cospectrum, the two PDSs, and a total PDS including the counts from both detectors, both in the zero-dead time and the dead time-affected cases. 
We then fitted the resulting spectra with a Lorentzian model in XSPEC.

\fref{fig:freq} shows the change of the \rms measured with the PDS and the cospectrum, with and without dead time.
The measured \rms in the  zero-dead time PDSs agrees with the zero-dead time cospectrum, whereas the dead time-affected cospectrum yields a frequency-dependent deviation from the true \rms, with deviation following the same trend as the variance (see also \fref{fig:cpdsvspds}).

The detection significance does {\em not} depend on the frequency in any case, with or without dead time. 
The decrease of the significance is instead driven by the observed count rate, as we discuss shortly.
In the no-dead time case, the significance of the single PDSs is about half that of the total PDS because the significance is directly proportional to the intensity of the signal (\eref{eq:sig}), and in the total PDS one uses twice the number of photons.
The zero-dead time cospectrum yields instead a significance $\sim\sqrt{2}$ lower than the total PDS, and higher by the same amount than the single-module PDS. 
This is just an effect of the factor $\sqrt{2}$ between the standard deviations of the cospectrum and the single-module PDS.
From \fref{fig:freq}, it is clear that it is advantageous to use the total PDS for low count rates where dead time is negligible, and the cospectrum otherwise, but with the formulas and the simulations shown above to account for the frequency-dependent distortion of the \rms amplitude.

In summary, the important point that  \fref{fig:freq}  makes is that QPOs are still detectable at any frequency, even those heavily affected by dead time, albeit with a change of the measured \rms that must be taken into account.

\paragraph{Count rate dependence}
We now investigate how the measured \rms is influenced by count rate.
\fref{fig:cts} and~\ref{fig:ctsback} show the variation with count rate in the detected \rms of a QPO at 30\,Hz, FWHM=2\,Hz and \rms=7.5\%, in two cases: increasing total count rate, and fixed total (source + background) count rate with variable source count rate.
For the first case, since $30\,{\rm Hz}<<1/\deadt$ and the higher-order corrections are not needed, we use van der Klis (1989; equations 3.8 and 4.8) to obtain
\begin{equation}\label{eq:drms}
\frac{\rms_{\rm det}}{\rms_{\rm in}} \approx \frac{1}{1+\deadt\rin} = \frac{\rdet}{\rin}.
\end{equation}
This relation is plotted with a dashed line in the top panel of \fref{fig:cts} and it is in remarkably good agreement with the simulated data.

In general, one would expect the significance of detection in the PDS to be proportional to the incident count rate and to the square of the \rms (\eref{eq:sig}). 
This condition holds if the QPO can be considered a small disturbance in an otherwise Poissonian process, or $\rms << \sqrt{\nu/\rin}$ \citep{Lewin+88}. 
In the bottom panel of \fref{fig:cts}, we fit \eref{eq:sig} below 600\,\cts, with a multiplicative constant due to the slightly different definition of 1-$\sigma$ error that we use ($\Delta\chi^2=1$ instead of the \citet{Leahy+83} definition).
The best-fit multiplicative constant, $\sim1./2.2$, turns out to be consistent with the factor 2 expected from the fact that we are using Gaussian fitting instead of excess power (see \sref{sec:simutec}).
The departure from the linear condition above $\sim600$\,cts is evident.
Indeed, it is expected that at count rates above $0.1~\nu/\rms$, the significance starts departing from the linear trend.
The total PDS is visibly more affected because its count rate is double than that of the single-module PDS.
The significance of detection with the cospectrum is $\sim\sqrt{2}$ lower than that of the total PDS, while it is $\sim\sqrt{2}$ higher than that of the PDS from a single module. 
The dead time-affected cospectrum, instead, has a large deviation from the linear trend.
This is just an effect of the diminishing count rate due to dead time.
In fact, what is plotted is the {\em incident } count rate.
If one converts it to the {\em detected} count rate, the linear relation between count rate and significance still holds (solid line).

The second case (\fref{fig:ctsback}) clearly shows a linear decrease of the measured \rms as the source gains counts with respect to the background.
Again, by using van der Klis (1989; equations 3.8 and 4.8), but this time putting the {\em total} count rate ($\rin + r_{\rm back, in}$, where $r_{\rm back, in}$ is the non-source count rate) in the relation between incident and observed count rates, one obtains
\begin{equation}\label{eq:drmstot}
\frac{\rms_{\rm det}}{\rms_{\rm in}} \approx \left[ 1 - \frac{\deadt\rin}{1+\deadt(\rin + r_{\rm back, in})} \right]
\end{equation}
This means that the measure of \rms we obtain in our data will be generally affected more if the source signal dominates the background, as is the case in most \nus observations of bright sources. 
In the examples that we present in the next Sections, we make use of Monte Carlo simulations similar to the ones above to estimate the change of \rms at the count rate of the sources we observe.

\section{Observations and data reduction}\label{sec:red}
\begin{table}
\centering
\caption{Summary of the observations used in this work}
\label{tab:obs}
\scriptsize {
\begin{tabular}{lcccc}
\hline
\hline
Source & ObsId & Date      & On time & Livetime \\
       &       &  y-m-d &   ks    &   ks\\
\hline
\grs & 10002004001 & 12-07-03 & 27.0 & 15.7 \\
\cyg & 30001011002 & 12-10-31 & 18.4 & 10.8 \\
\cyg & 30001011003 & 12-10-31 & 10.3 & 5.1 \\
\gx & 80001013002 & 13-08-12 & 38.8 & 35.1 \\
\gx & 80001013004 & 13-08-16 & 23.4 & 20.8 \\
\gx & 80001013006 & 13-08-24 & 26.8 & 23.2 \\
\gx & 80001013008 & 13-09-03 & 11.1 & 9.4 \\
\hline
\end{tabular}
}
\end{table}
We now demonstrate the methods outlined above by analyzing the power density spectra for a number of bright Galactic black hole binaries.

A summary of the observations used in this work is listed in \tref{tab:obs}.   
We preprocessed all \nus observations with the NUSTARDAS pipeline included in HEASOFT 6.15.
We produced cleaned event files and calculated good time intervals (GTIs).
We processed both the unfiltered and the clean event files (files tagged with ``\_uf'' and ``\_cl'' respectively in the ``event\_cl'' directory of \nus data). 
By analyzing the unfiltered and cleaned event files, it is possible to estimate the ratio of ``good'' events to the total of dead time-producing events. 

We calculated the ratio $r_{\rm good}$ of ``good'' events over intervals of 100s.
We found that in all observations $r_{\rm good}$  was almost constantly $0.9 \lesssim r_{\rm good}\lesssim 0.94$, with some drops due to increased solar activity.
We selected the intervals in which this ratio didn't go below a certain threshold (ranging from 0.9 to 0.92 in different files), in order to eliminate possible spurious effects on the observed variability. 
In the subsequent analysis, we will always consider an additional 10\% ``background'' event rate due to unrecorded and spurious events.

We selected events from a region around each source depending on how broad the point-spread distribution was (see the next subsections for details for each source).

We used the {\tt barycorr} FTOOL to calculate event times at the solar system barycenter using the DE200 ephemeris and source positions from Simbad\footnote{{\tt http://simbad.u-strasbg.fr/}}. 

Additional variability is introduced into the cross-power spectrum in some cases due to mis-matched GTI between FPMA and FPMB. This mis-match occurs often when the satellite traverses the South Atlantic Anomaly (SAA) because the automatic algorithm that turns off the focal plane modules to conserve telemetry kicks in at different thresholds for FPMA and FPMB.
The parameters that rule this behavior are finely tuned in order to maximize the number of photons available for spectral analysis while avoiding most of the contamination \citep{nustar13}, but for timing analysis this produces unwanted effects that have to be accounted for.
A signature of this variability is a strong time lag between the two detectors in the same energy band 
(therefore, unrelated to the physics of the source), that disappears when discarding an additional interval at the GTI borders or by using a more aggressive filtering during the pipeline process.
We decided to use a ``safe'' additional cut of 300s at the borders of all GTIs during timing analysis, and verified that it solved the problem in all affected ObsIds.

\section{Data analysis}\label{sec:analysis}
From each barycentered event list (\sref{sec:red}) we obtained light curves sampled at $1/2048$\,s ($\sim488\mu$s).
We extracted separate light curves for each detector, and calculated the common Good Time Intervals (GTIs). 

Each light curve was then divided in segments (the length was chosen in order to optimize the use of GTIs) with the start and stop times locked between the two detectors,
and a Fourier transform was calculated from each of them.
We then calculated the CPDSs from each segment by multiplying the Fourier transforms as explained in \sref{sec:deadtime},
and obtained a single cospectrum from the real part of the average of all CPDSs. 

When needed, we plot the error on the best-fit model as a hatched region around the model line.
To calculate these errors, we ran a Monte Carlo Markov Chain on XSPEC to simulate 10000 models around the best fit.
We chose all the models having $\Delta\chi^2<2.706$, and we calculated at each energy the maximum deviation from the best fit coming from any of these models.


\subsection{\cyg}\label{sec:anacyg}

\cyg is a bright persistent BHB, discovered during one of the earliest rocket flights at the dawn of X-ray astronomy \citep{Bowyer+65cyg}. 
As such, it is among the best studied X-ray sources in the Galaxy. 
Its spectral evolution is typical for a BH, with the succession of low/hard, high/soft and intermediate states with correspondingly different timing patterns \citep[see][for a broad view on BHBs]{Dunn+10}.
Variability in all spectral states has been observed.
It shows two dominant Lorentzian components in the low/hard state.
When the source softens their characteristic frequencies increase and an
additional cutoff power law-like component appears. In the soft state
the latter dominates the power spectrum \citep[and references therein]{Nowak+96,Nowak+99, Revnivtsev+00, Churazov+01, Axelsson+05,Boeck+11}.
Time lags have been observed in all spectral states \citep{Pottschmidt+00}; the variability of single spectral components was also investigated, showing that the reflection component shows variability at lower frequencies than the bulk of the emission in the low/hard state \citep{Revnivtsev+99} while the \rms level and shape of the power spectrum is comparable at all energies in the soft state \citep{Gilfanov+00}. 
A very thorough investigation of the relation between variability in different energy ranges and spectral states can be found in \citet{Grinberg+14}. For past studies of the high-energy variability of the source see \citet{Cabanac+11}.

\nus observed \cyg in its soft state \citep[for a full spectral analysis description, see][]{Tomsick+14}.
The spectral analysis shows a strong reflection component, that might produce detectable phase lags. 
Phase lags have been observed with \rxte in all spectral states (\citealt{Miyamoto+89, Pottschmidt+00,Boeck+11,Grinberg+14}; see \citealt{Nowak+96,Nowak+99} for a rigorous treatment).

\begin{figure*}
\centering
\includegraphics[width=\columnwidth]{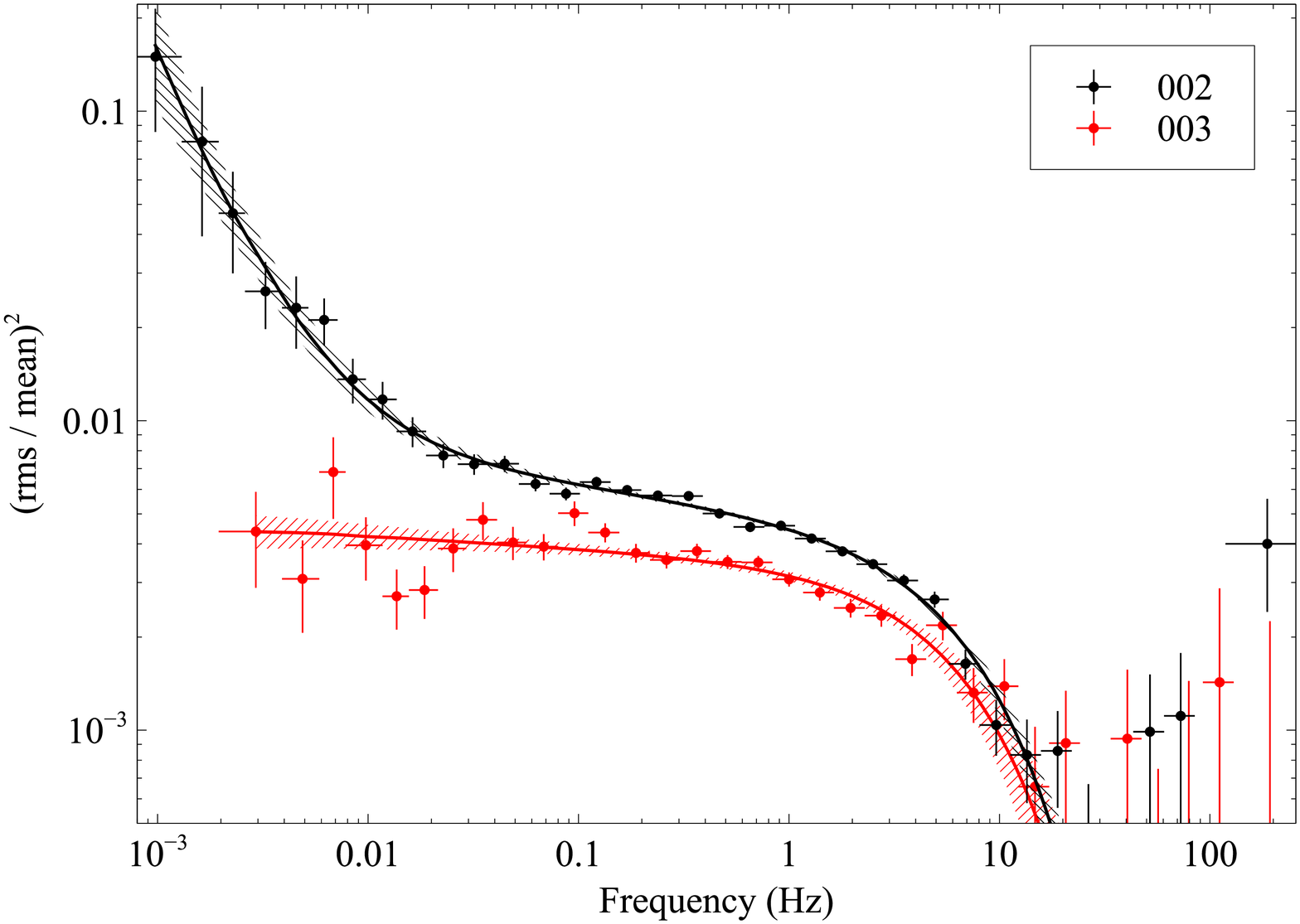}
\includegraphics[width=\columnwidth]{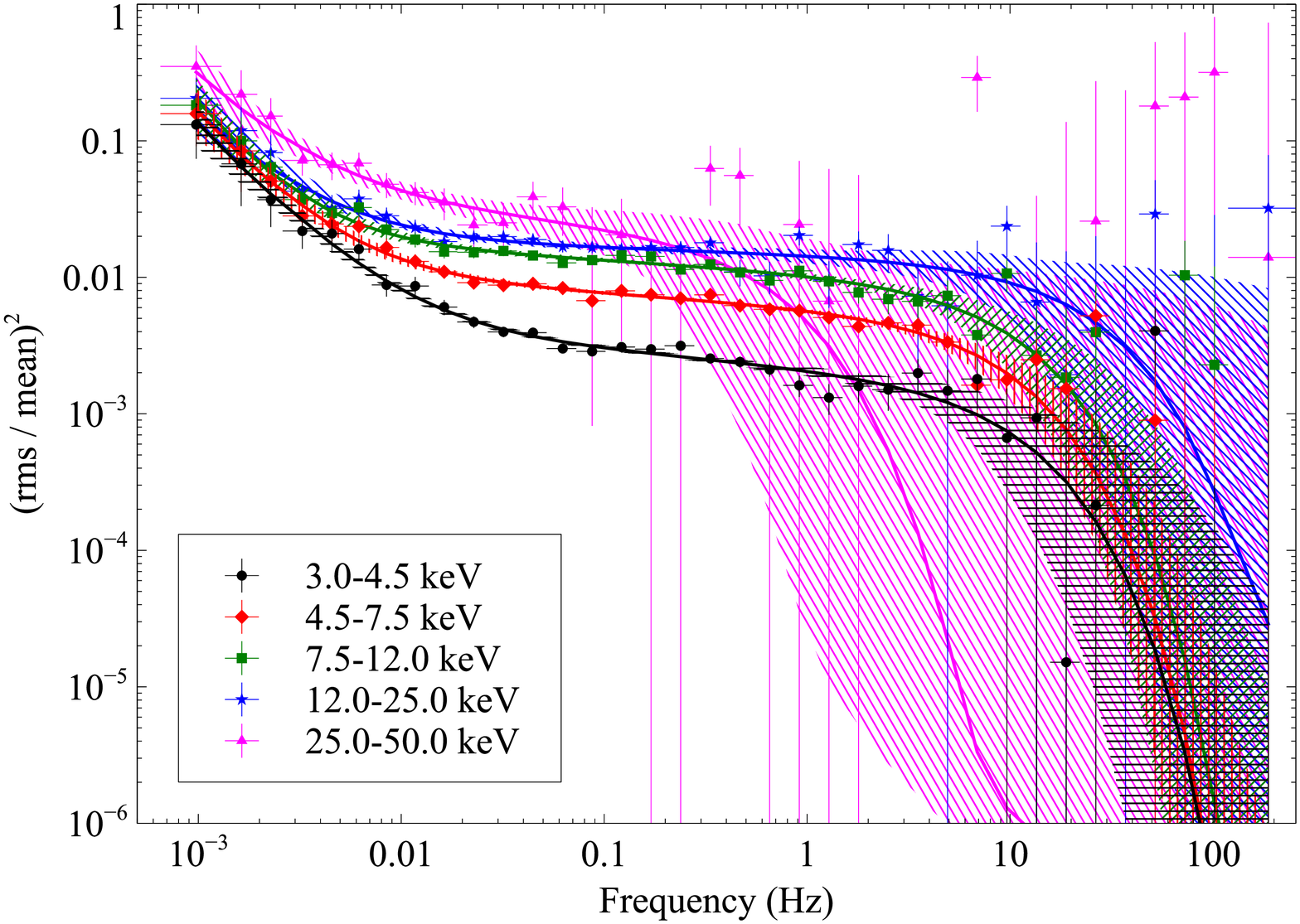}
\includegraphics[width=\columnwidth]{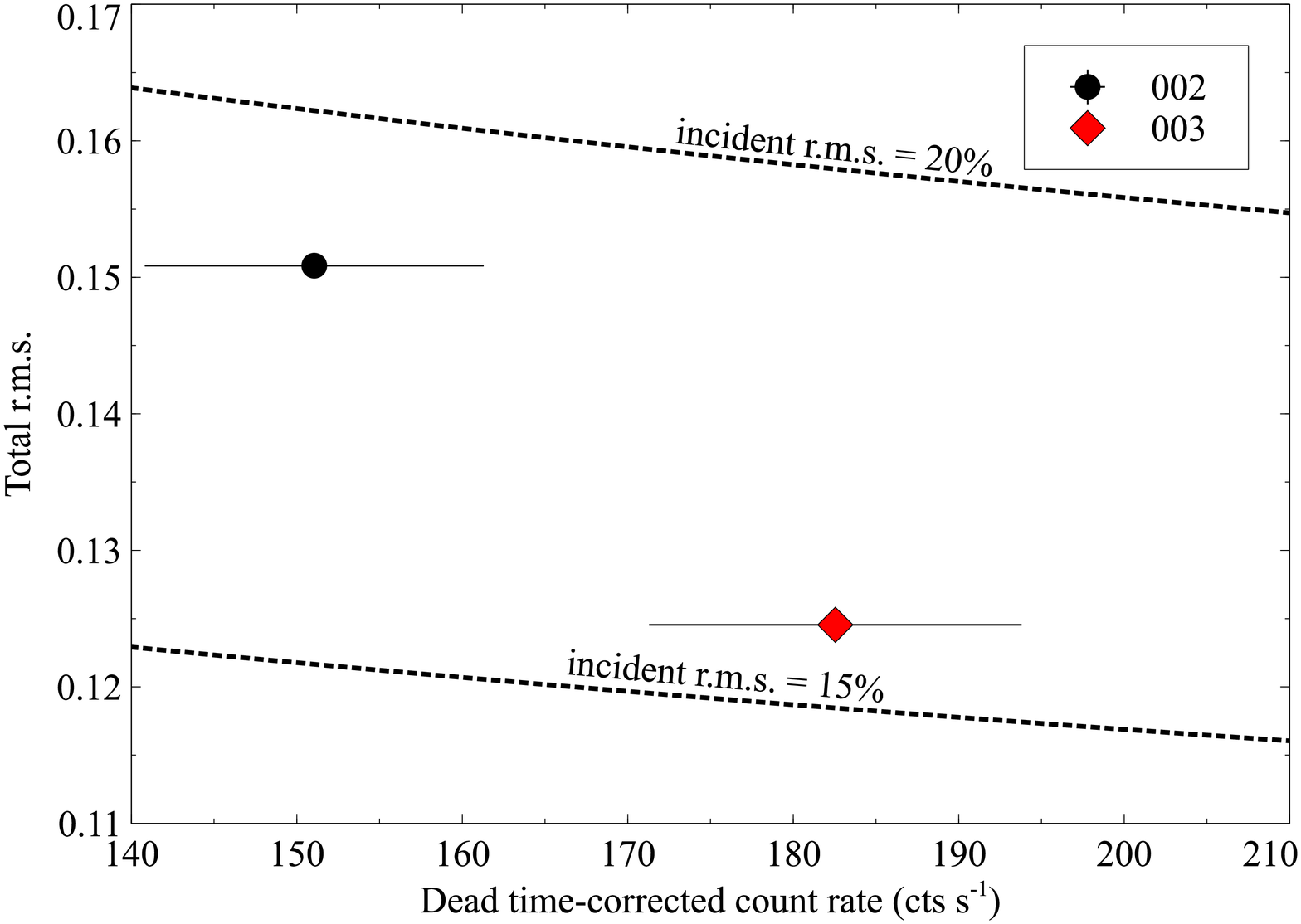}
\includegraphics[width=\columnwidth]{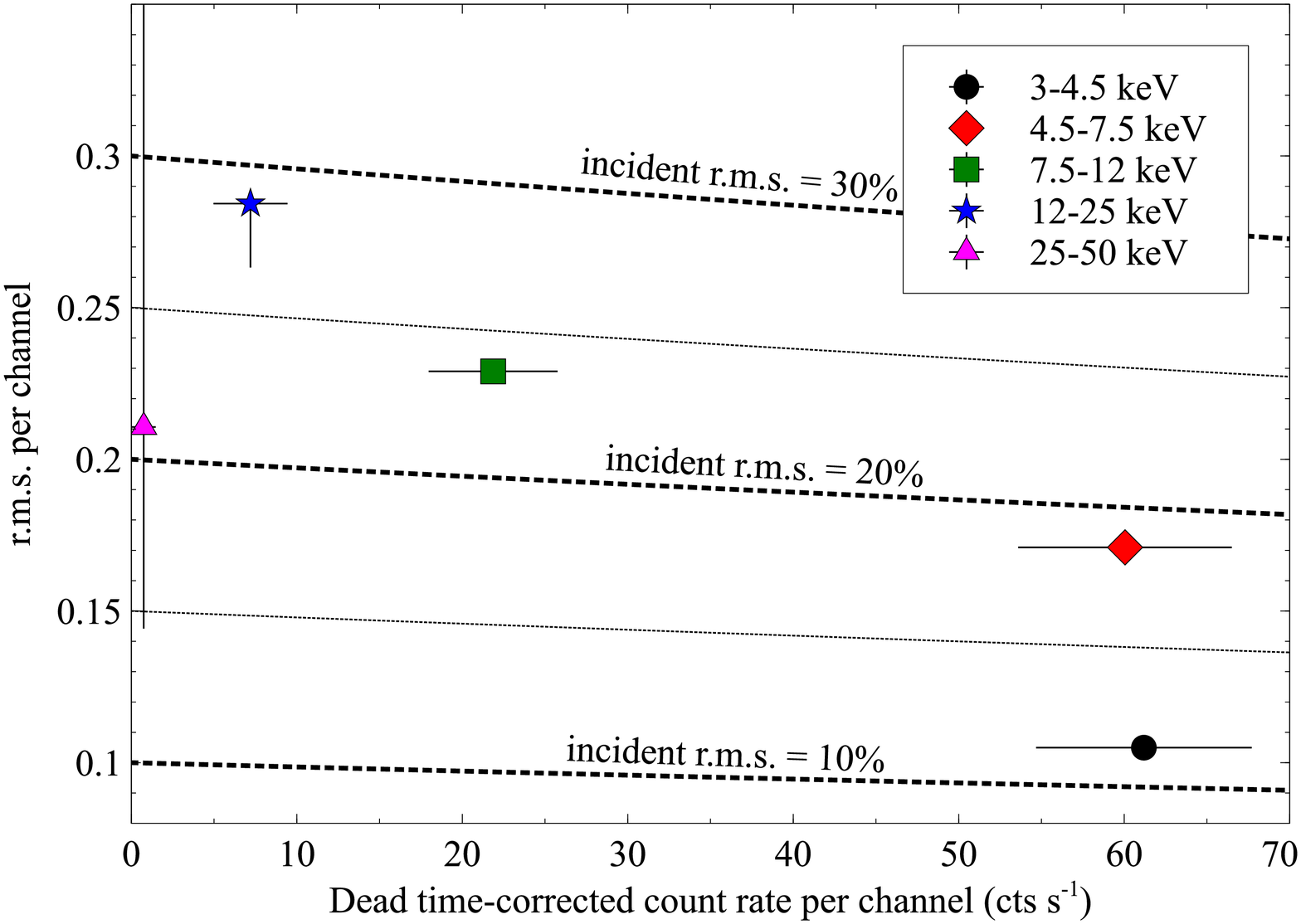} 
\caption{\cyg: ({Top Left}) Comparison of the cospectrum in the two ObsIds; the first observation is fitted with a {\tt powerlaw+cutoffpl} model to account for the low-frequency excess, while the second with a simple {\tt cutoffpl} model (the power law was not required by the fit). 
({Top Right}) Cospectra in different energy ranges. There is a clear change of the cospectrum with energy, with a monotonic increase of \rms. The cutoff frequency is consistent with being stable. 
Hatched regions represent the 90\% errors ($\Delta\chi^2=2.706$) on the best-fit {\tt powerlaw+cutoffpl} model for each energy range. 
(Bottom Left) Change of total \rms in \cyg between the two observations. The points give the measured \rms, and the dashed lines show a grid of dead time-corrected \rms values, following the methods shown in \fref{fig:cts}.
(Bottom Right) Change of \rms at different energies in obs. 002.
The \rms error bars are smaller than the marker size in most cases.}
\label{fig:cyg_pdsvse}\label{fig:cyg:rms}
\end{figure*}

The PDS in this state is often well described by a cutoff power law-like shape almost ``flat'' between $10^{-2}$ and $10$\,Hz  \citep[e.g.][]{Cui+97,Churazov+01,Gleissner+04,Axelsson+05}.
This is the case in our observations, as shown in \fref{fig:cyg_pdsvse}.
There is a clear change in \rms variability between the two observations, and a power law-like red noise component appears at the lowest Fourier frequencies in only one observation, but for the rest the cospectrum is very well described by a cutoff power law.

As shown in \fref{fig:cts}, a drop of \rms is expected at higher count rates, and the second observation has indeed a higher count rate than the first.
But in \fref{fig:cyg:rms} we show that the expected \rms drop due to dead time is far lower than observed.
This means that the rms change is largely due to an intrinsic lower variability of the source in the second observation.

More interesting is the cospectrum in different energy ranges, shown in \fref{fig:cyg_pdsvse}.
There is a very strong change in the shape and normalization of the cospectrum with energy.
In general, the \rms increases with energy, and it is about an order of magnitude higher above 10 keV than below 5 keV.
This is a clear indication that the soft component of the spectrum (coming from the disk) is varying less than the reprocessed component (reflected and Comptonized).

Again, in order to evaluate the \rms change with energy, it is necessary to apply a correction for dead time.
In fact, channels with more counts will have a larger drop of \rms than channels with a few counts (compare to \fref{fig:ctsback}).

Following the procedure in \sref{sec:simu}, we simulated the expected relative drop in \rms in all bands.
In \fref{fig:cyg:rms} we show the measured \rms and the curves showing the expected drop due to count rate.
These curves were calculated with the method depicted in \fref{fig:cts}, assuming a total count rate $\sim1.1$ times the source count rate, as the total background plus the rejected events account for about 10\% of the contributions to dead time.
As can be seen, the \rms change between bands is mostly intrinsic, with a drop of only $\sim10$\% expected for the channels with the highest count rates.
Our cospectrum results are consistent with the PDS results from other works \citep[e.g.][]{Grinberg+14}.

Although we do not find any significant time lags between any of the energy bands, this is mostly due to the fact that our observation is not long enough. 
We estimate that $\sim40$\,ks should be sufficient to detect time lags similar to those observed in the past in \cyg, in this spectral state \citep[e.g.][]{Pottschmidt+00}.

\subsection{\gx}\label{sec:anagx}

\gx is another well known BHB.
Discovered in 1973 by {\em OSO-7} \citep{Markert+73}, it has been observed frequently since then. 
It is a transient system, with long outbursts and periods of quiescence lasting up to several years \citep[and references therein]{Belloni+05}.
Monitored during its outbursts by \rxte, it showed one of the best examples of q-track pattern in the hardness-intensity diagram (\citealt{Belloni+05}; for the q-track pattern see \citealt{MaccaroneCoppi03} and \citealt{Dunn+10} for a comparison with other sources).
Like many accreting BHs, its low/hard state PDS is dominated by two or more Lorentzian components, whose frequencies are positively correlated to count rate. 
As the source reaches the intermediate and soft state, the broad Lorentzians move further to higher frequencies and various kinds of QPOs appear \citep[see, e.g.,][]{Belloni+05,Casella+05}, including high frequency QPOs.
Thanks to this wealth of data, it has often been used, like \cyg, to test new timing approaches.
In its very high state, the signal at low energy ranges is generally coherent, while coherence is lost between low and high energies \citep{Vaughan+97}.
\citet{Nowak+99339} measured coherence and lags in the hard state of the source, showing that the two Lorentzians are likely produced by independent, but internally coherent, processes.
\citet{Cassatella+12} showed that hard time lags in this source are more likely to be produced by propagation of modulations of the accretion flow in the disk rather than from reflection.
 
\nus observed \gx four times during the rise of the last outburst.
During this outburst, the source remained in a canonical low/hard state without reaching the threshold luminosity that would have marked the transition to the intermediate and soft states.
The spectral analysis will be published in a separate publication (F\"urst et al 2014, \inprep), but in this Paper we will concentrate on the timing properties of the source, analyzed with the methods described above.

\begin{figure*}
\centering
\includegraphics[width=\columnwidth]{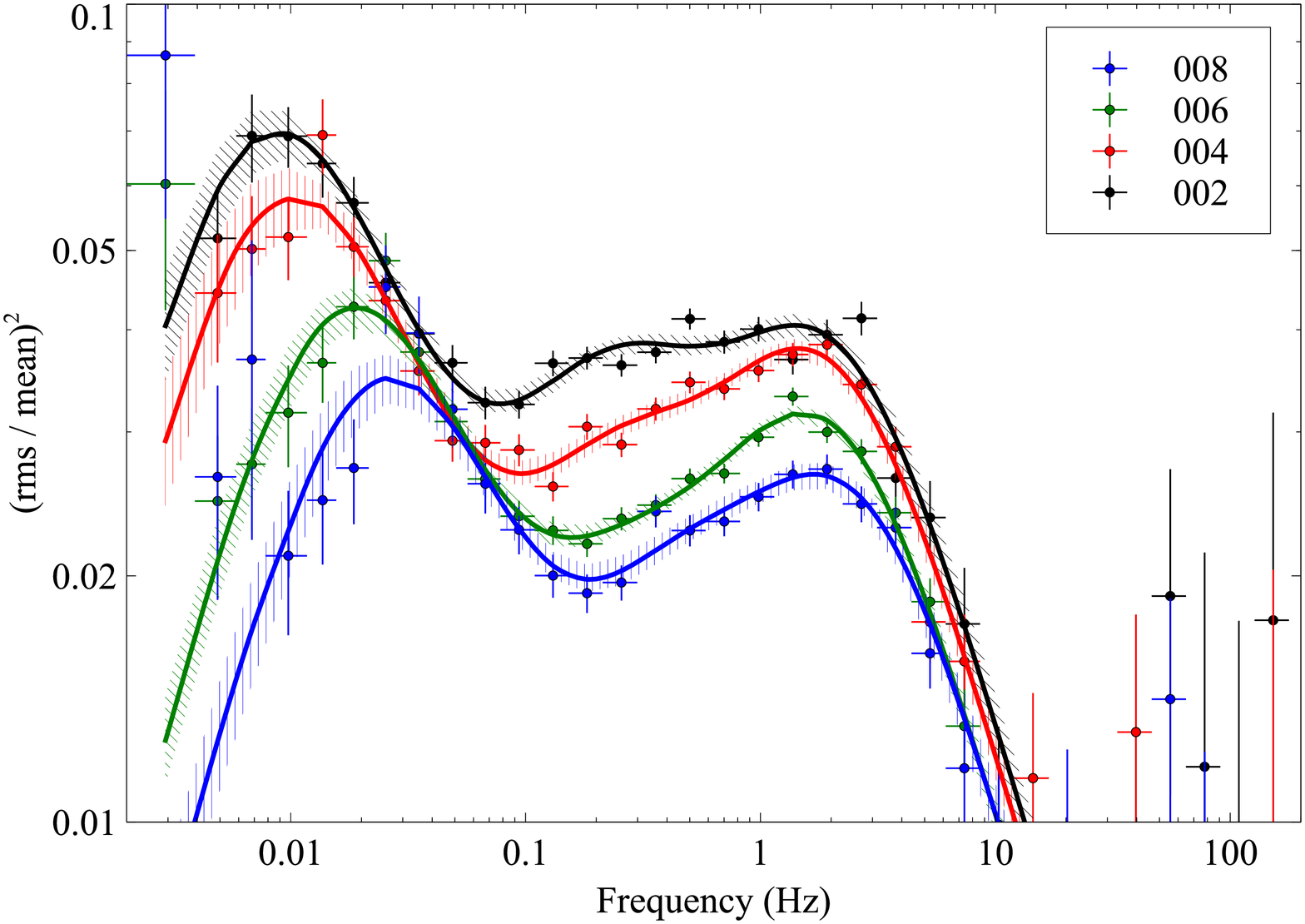}
\includegraphics[width=\columnwidth]{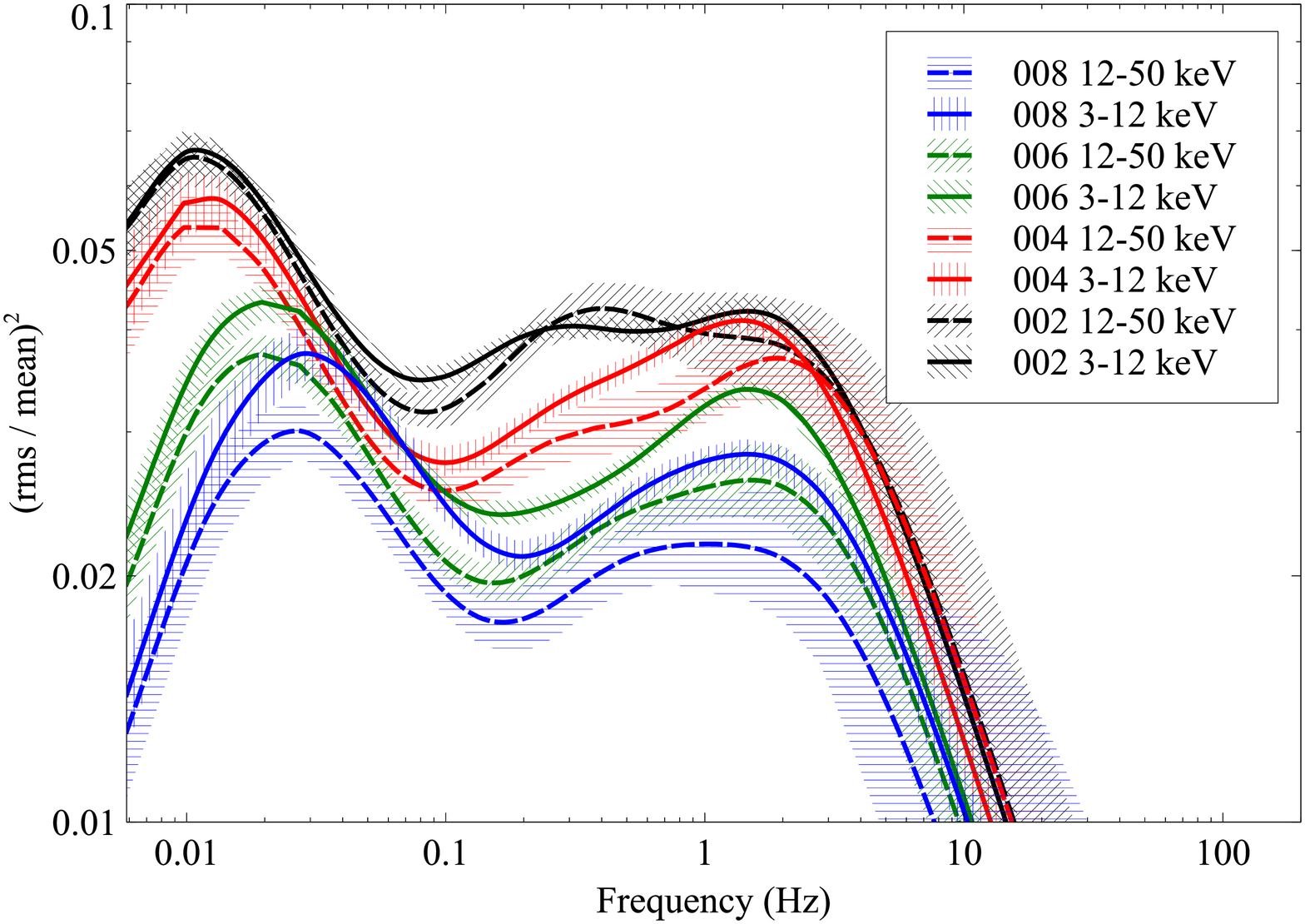}
\caption{\gx: 
Cospectra for the four pointings, in the 3--50 energy band (Left) and in two energy bands (Right).
Best-fit models with 3 zero-centered Lorentzian components are also plotted.
In the second plot, only the best fit models are plotted for clarity.
There is a clear evolution of the cospectrum: first of all, the \rms generally decreases as the outburst goes on; the Lorentzians, in particular below 1Hz, generally increase their widths; also, while the two energy bands have the same cospectrum in the first observation, they separate more and more clearly in the following ones.
We verified that this change is too strong to be produced by dead time alone (it can account for at most a 10\% relative \rms drop).
In the latest two observations a hint of a power law component below 0.01\,Hz also appears. It is not significant in our fits.}
\label{fig:gx_4obs}
\end{figure*}

The whole band and energy-dependent cospectra in the four different observations are shown in \fref{fig:gx_4obs}.
The approximate flux of the source increased almost linearly from $\sim10$ to $\sim40\,\cts$ in each detector between the first and the last observation.
The general shape and the variation of the cospectrum is similar to what has been observed in the past for this source in the hard state \citep[e.g.][]{Nowak+99339,Belloni+05} and for other BHBs \citep{McClintockRemillard06}, with two zero-centered Lorentzian components dominating the shape (L$_1$ and L$_2$ in \citealt{Belloni+05}), plus occasionally a third Lorentzian needed to account for an excess between them.

The cospectrum evolves with L$_1$ consistently increasing its characteristic frequency in subsequent observations (as luminosity increases), while L$_2$ does not change as much.
This is consistent with what was previously reported by \citet{Belloni+05} using the PDS.
The energy dependence of the cospectrum is very weak in the first observation, and tends to increase in the subsequent observations, while the total \rms consistently decreases as the count rate increases.
In the second, third and fourth observation the high energy cospectrum has consistently a lower \rms than the low energy one.
Again, these cospectra results are consistent with previous PDS results \citep[e.g.][]{Nowak+99339}.

From \fref{fig:cts}, and the discussion about the much brighter \cyg, it is clear that at these frequencies and for count rates changing from 10 to 40\,\cts we do not expect a such a strong change in \rms, which must therefore be intrinsic to the source.

A more detailed discussion will be presented in F\"urst et al., \inprep.

\section{Conclusions}\label{sec:concl}
This paper presents a method to exploit \nus's two independent detectors to work around the spurious effects 
resulting from dead time, enabling standard aperiodic timing analysis of X-ray binaries to be applied.
As extensively studied in literature \citep[e.g.][]{VDK89,Vikhlinin+94,Zhang+95}, dead time produces a spurious correlation between event times that strongly modifies the shape of the PDS. 
For frequencies $\nu << 1/\deadt$, the only visible effect in PDSs is a slight decrease of the white noise level. 
In timing-oriented X-ray missions, dead time is very short (\rxte/PCA: 10\,$\mu$s, \citealt{Jahoda+06}; {\em ASTROSAT}/LAXPC: 10--35\,$\mu$s, \citealt{Paul09}; {\em LOFT}: $<1$\,ns, \citealt{Feroci+12}) and the ``interesting'' range of frequencies for accretion (the highest being the dynamical timescales around neutron stars $\simeq 2000$\,Hz) is well below $1/\deadt$.
In \nus, \deadt is around 2.5\,ms, right in the middle of the frequency range commonly analyzed in BH and NS binaries. 
Moreover, in order to obtain better precision in the measure of photon energies, the dead time is variable and depends on the event grades.
Due to this, standard techniques for the modeling of the PDS shape \citep[e.g.][]{Vikhlinin+94,Zhang+95} are not applicable.

But \nus has two independent focal plane modules that are read out by separate processors,
and the dead time is therefore independent.
If, instead of taking the PDS of each detector, we take the cospectrum (see \sref{sec:cpds}) of the signals in the two detectors, one is able to cancel out most correlations produced by dead time and obtain a close proxy of a white noise-subtracted PDS.

We described in Sections~\ref{sec:cpds} and~\ref{sec:simu} how to calculate the cospectrum and estimate the remaining effects of dead time on the signal,
namely the modulation of the variance of data bins,
that cannot be described using the standard \citet{Leahy+83} prescriptions,
and the equivalent change of the observed \rms at different frequencies and for different count rates.
This estimate requires Monte Carlo simulations tailored to the source that is being analyzed, as described in \sref{sec:simu}.

We also applied this technique to the analysis of \nus data of two BHBs, \cyg and \gx. 
\cyg was observed in the soft state.
We report a strong change of \rms with energy, with a significant rise of the variability at higher energies. 
These cospectrum results are consistent with earlier PDS results, and extend the energy coverage with respect to earlier results from \rxte \citep[e.g.]{Grinberg+14}. 
Comparing these results with the spectral modeling presented by \citet{Tomsick+14}, it is apparent that the low energies are dominated by the disk emission, while at higher energies it's the reprocessed component (Comptonized and reflected) that dominates.
This is consistent with what is found in many BHBs, and in particular to what has already been reported in this source \citep[the ``stable disk and unstable corona'' picture, e.g. ][]{Churazov+01}.

In \gx, we observed the rising part of the outburst, in a standard low/hard state.
The cospectrum shows that, as the outburst goes on, the \rms generally decreases and the characteristic frequency of the first Lorentzian component used to described the cospectrum consistently rises.
Also, this is consistent with what has been observed in the low/hard state of this and other sources using the PDS \citep{Nowak+99339,Belloni+05}.

We do not find evidence for time lags in the sources analyzed in this paper, due to the short observing time available, but the cross power density spectrum can be used to calculate time lags without major impact from dead time, if data are adequately filtered to avoid intervals with high background activity and in particular at the start and end of each good time interval.

The techniques presented in this paper can be used to perform most of the standard timing and spectral timing analysis used in literature, if the change of \rms with frequency and count rate is properly taken into account (\sref{sec:simu}). 
For slow timing ($\nu<<1Hz$), the dead time correction included in the official pipeline is still recommended, because it effectively corrects for all sources of dead time, including shielded events. 
For low count rates ($\lesssim 1\cts$) the standard PDS of the sum of the two lightcurves (what we called the ``total PDS'' above) is still advantageous, because the effects of dead time are negligible and the sensitivity of the total PDS, expressed as significance of detection of a standard variability feature, is $\sim\sqrt{2}$ higher than the cospectrum.

As an additional note, this technique can be used in any satellite that has completely independent detectors, and of course by using multiple satellites. 
\rxte, with its independent PCUs, would be a good candidate if not for the presence of so-called Very Large Events (VLE), particle events depositing more than 100\,keV and occurring many times per second \citep[e.g.][]{Jones+08}.
These events often involve more than one PCU, invalidating the assumption of independence. 
The Large Area Xenon Propotional Counter (LAXPC) onboard the upcoming {\em ASTROSAT} mission \citep{astrosat06} will provide three independent detectors. 
Provided that the fraction of VLE-equivalent events is negligible, this technique will be applicable to LAXPC data.

\acknowledgements
MB and DB wish to acknowledge the support from the Centre National d'\'Etudes Spatiales (CNES) and the Centre National de la Recherche Scientifique (CNRS).
CS acknowledges funding by the German BMWi under DLR grant numbers 50~QR~0801, 50~QR~0903, and 50~OO~1111.
P.G. thanks STFC for support (grant reference ST/J003697/1).
This work was supported under NASA Contract No. NNG08FD60C, and
made use of data from the {\it NuSTAR} mission, a project led by
the California Institute of Technology, managed by the Jet Propulsion
Laboratory, and funded by the National Aeronautics and Space
Administration. We thank the {\it NuSTAR} Operations, Software and
Calibration teams for support with the execution and analysis of
these observations.  This research has made use of the {\it NuSTAR}
Data Analysis Software (NuSTARDAS) jointly developed by the ASI
Science Data Center (ASDC, Italy) and the California Institute of
Technology (USA). 
The timing analysis was executed with Matteo Bachetti's Libraries and Tools in Python 
for \nus Timing (MaLT PyNT).
This code is available upon request.
Most of the plots were produced with the Veusz software by Jeremy Sanders.
The authors wish to thank Chris Done, Denis Leahy and Tomaso Belloni for very insightful discussions.

\appendix

\section{\grs: timing analysis from instrument commissioning}\label{sec:anagrs}
\begin{figure*}
\centering
\includegraphics[width=0.48\linewidth]{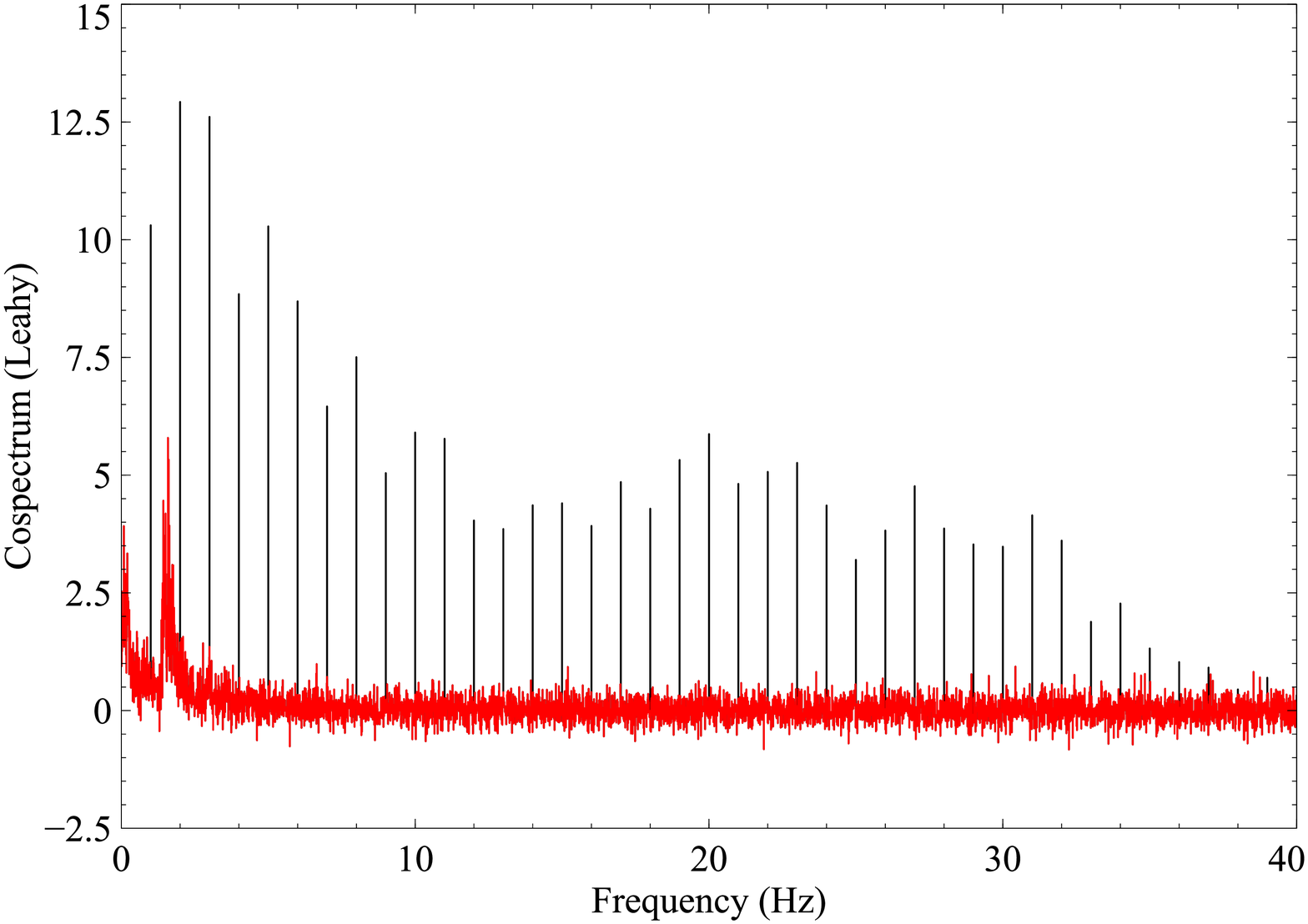}
\includegraphics[width=0.48\linewidth]{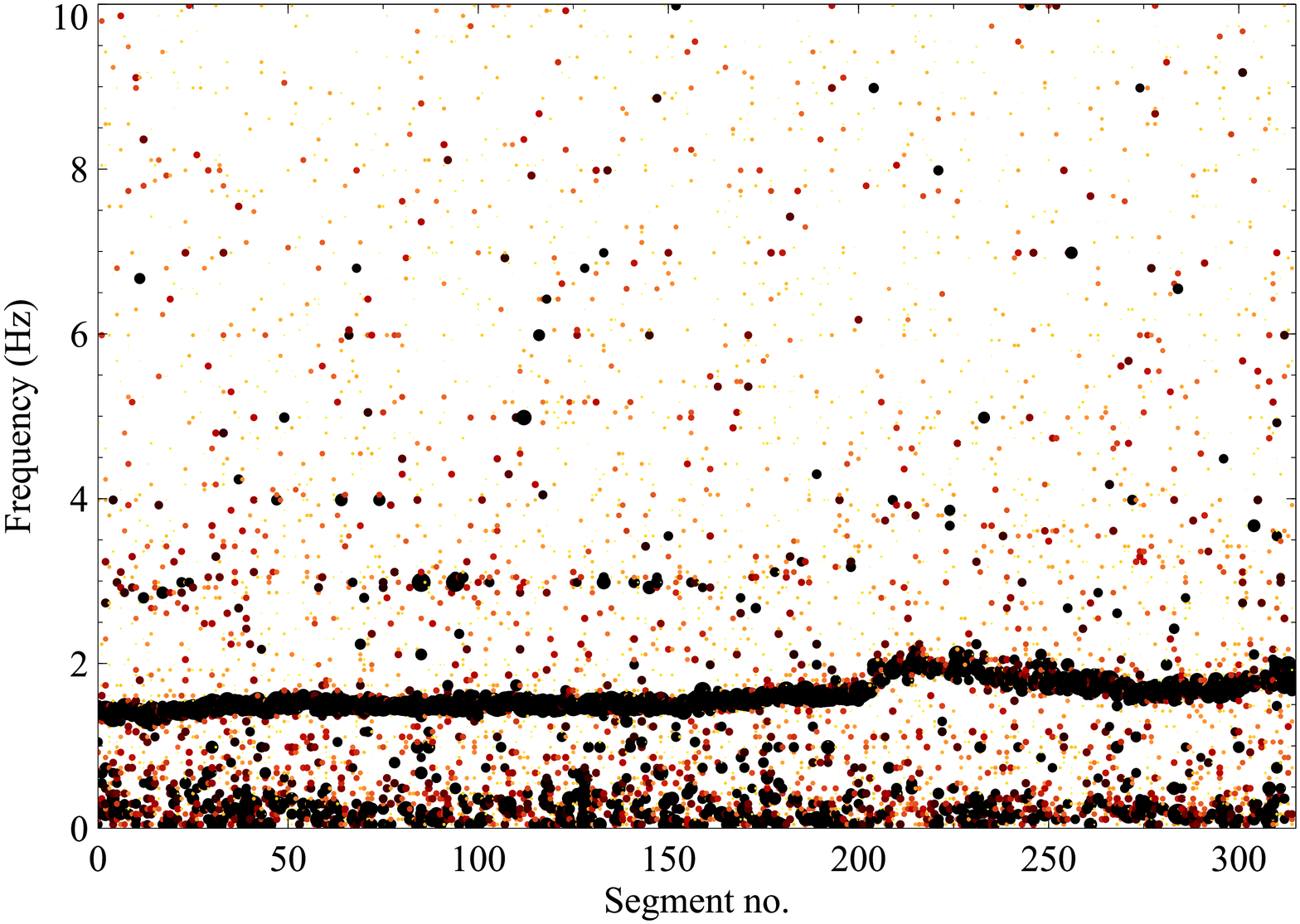}
\caption{\grs: 
(Left) Cospectrum before (black) and after (red) clean up of the nth harmonics of the housekeeping operations (at 1 and 0.125 Hz)". (Right) evolution of the 1.5\,Hz QPO (and a faint first harmonic) in the dynamical cospectrum during the observations. Each time segment represents 64s of gap-free data. Colors and sizes of the points are both a proxy of power, in arbitrary units.}
\label{fig:grscpds}
\end{figure*}

\grs is a very bright persistent microquasar, discovered in 1992 by {\em Granat} \citep{CastroTirado+92} which shows accretion states that depart from the usual behavior of Galactic BHs, being associated with very high-Eddington fractions. 
The variability of this source has been extensively studied in the past, in particular for the presence of very strong low frequency ($<$0.1\,Hz), medium-low frequency (0.5--10\,Hz), medium-high frequency ($\sim$40 and 67\,Hz) and high frequency (113 and 165\,Hz) QPOs \citep[e.g.][]{Morgan+97,Strohmayer01, Remillard+03, Pahari+13}.
In our observation, the medium-low QPO is the only one observed to very high significance. Its behavior has been studied in great detail \citep[e.g.][]{Reig+00} and its properties are known up to very high energies thanks to \rxte/HEXTE observations \citep{TomsickKaaret01}.
This source is discussed as a reference for timing analysis on sources observed very early in the mission.
The first version of the flight software, in fact, introduced a large number of spurious modulations in the signal that we will discuss and show a way to work around.

During the observatory commissioning phase prior to commencement of science observations, \nus observed the galactic BHB \grs in its {\em plateau} state \citep{Foster+96}.
The spectral analysis of that observation was presented by \cite{Miller+131915}.
The version of the flight software that was installed at that time on \nus executed housekeeping operations at regular intervals (1 or 8s), producing periodic dead time. 
Therefore, the raw power spectrum and the CPDS of this observation are plagued by spurious features at 1Hz and 0.125Hz as well as their harmonics and sub-harmonics (see \fref{fig:grscpds}).
In this Section we show a technique to clean up the cospectrum and perform a basic analysis of the strong QPO that this source.

To ``clean'' the cospectrum from these spurious features, we first produced a cospectrum on a time scale $t_{\rm fft}$ of 128\,s.
The maintenance operations were very regular, and so in this cospectrum the corresponding features were very sharp and limited to a single bin each.  
We replaced the value of each of these bins with a random value calculated from the mean of the two closest bins and their standard deviation, as shown in \fref{fig:grscpds}.
The possible spurious effects of this imperfect statistical treatment of the simulated bin value are negligible due to the small frequency bin (simulated data are only one bin every $1/t_{\rm fft}=1/128$\,Hz) and the rebinning applied in the following steps of the analysis.
The real and imaginary part were treated independently.

From this cleaned CPDS we calculated the cospectrum. An example is shown in red in \fref{fig:grscpds}.
The cospectrum shows a very strong and variable QPO at $\sim1.5$\,Hz, with \rms=10(1)\% (from the best-fit Lorentzian amplitude, and corrected for the expected $\sim20$\% drop at $\sim100\,\cts$, \eref{eq:drms}). 
This feature has previously been observed in this spectral state \citep[e.g.][]{Pahari+13}.
In \fref{fig:grscpds} we show that with these techniques it is possible in principle to also apply techniques of QPO tracking (a dynamical cospectrum) given a sufficient \rms. 
A hint of the harmonic of this QPO is also observable, both in the single and in the dynamical cospectra.
No other significant QPOs are detected.
This source is known to show high-frequency QPOs in \rxte observations \citep{Morgan+97, Remillard+03}. 
These features are not observed in our dataset.
We calculate a rough upper limit of $\sim 5$\% for additional QPOs between 50 and 100\,Hz.
This upper limit is higher than the observed \rms of QPOs in this frequency range observed in the past (e.g. $\sim1.5$\% in the full \rxte band in \citealt{Morgan+97}), and our non-detection might be largely due to the low statistics of this dataset.


\begin{thebibliography}{76}
\expandafter\ifx\csname natexlab\endcsname\relax\def\natexlab#1{#1}\fi

\bibitem[{Agrawal(2006)}]{astrosat06}
Agrawal, P.~C. 2006, ASR, 38, 2989

\bibitem[{Arnaud(1996)}]{Arnaud96}
Arnaud, K.~A. 1996, ADASS, 101, 17

\bibitem[{Artigue {et~al.}(2013)Artigue, Barret, Lamb, Lo, \&
  Miller}]{Artigue+13}
Artigue, R., Barret, D., Lamb, F.~K., Lo, K.~H., \& Miller, M.~C. 2013, arXiv,
  1303.0248, 1303.0248

\bibitem[{Axelsson {et~al.}(2005)Axelsson, Borgonovo, \& Larsson}]{Axelsson+05}
Axelsson, M., Borgonovo, L., \& Larsson, S. 2005, A{\&}A, 438, 999

\bibitem[{Bachetti {et~al.}(2013)Bachetti, Rana, Walton, Barret, Harrison,
  Boggs, Christensen, Craig, Fabian, F{\"u}rst, Grefenstette, Hailey,
  Hornschemeier, Madsen, Miller, Ptak, Stern, Webb, \& Zhang}]{Bachetti+13}
Bachetti, M. {et~al.} 2013, ApJ, 778, 163

\bibitem[{Barret \& Vaughan(2012)}]{Barret+12}
Barret, D., \& Vaughan, S. 2012, ApJ, 746, 131

\bibitem[{Belloni \& Hasinger(1990)}]{BelloniHasinger90}
Belloni, T., \& Hasinger, G. 1990, A{\&}A, 230, 103

\bibitem[{Belloni {et~al.}(2005)Belloni, Homan, Casella, van~der Klis, Nespoli,
  Lewin, Miller, \& Mendez}]{Belloni+05}
Belloni, T., Homan, J., Casella, P., van~der Klis, M., Nespoli, E., Lewin, W.
  H.~G., Miller, J.~M., \& Mendez, M. 2005, A{\&}A, 440, 207

\bibitem[{Bendat \& Piersol(2011)}]{BendatPiersol11}
Bendat, J.~S., \& Piersol, A.~G. 2011, {Random Data: Analysis and Measurement
  Procedures} (Wiley)

\bibitem[{B{\"o}ck {et~al.}(2011)B{\"o}ck, Grinberg, Pottschmidt, Hanke, Nowak,
  Markoff, Uttley, Rodriguez, Pooley, Suchy, Rothschild, \& Wilms}]{Boeck+11}
B{\"o}ck, M. {et~al.} 2011, A{\&}A, 533, 8

\bibitem[{Boutelier(2009)}]{BoutelierThesis}
Boutelier, M. 2009, PhD thesis, PhD Thesis, CESR

\bibitem[{Boutelier {et~al.}(2009)Boutelier, Barret, \& Miller}]{Boutelier+09}
Boutelier, M., Barret, D., \& Miller, M.~C. 2009, MNRAS, 399, 1901

\bibitem[{Bowyer {et~al.}(1965)Bowyer, Byram, Chubb, \&
  Friedman}]{Bowyer+65cyg}
Bowyer, S., Byram, E.~T., Chubb, T.~A., \& Friedman, H. 1965, Science, 147, 394

\bibitem[{Cabanac {et~al.}(2011)Cabanac, Roques, \& Jourdain}]{Cabanac+11}
Cabanac, C., Roques, J.-P., \& Jourdain, E. 2011, ApJ, 739, 58

\bibitem[{Casella {et~al.}(2005)Casella, Belloni, \& Stella}]{Casella+05}
Casella, P., Belloni, T., \& Stella, L. 2005, ApJ, 629, 403

\bibitem[{Cassatella {et~al.}(2012)Cassatella, Uttley, Wilms, \&
  Poutanen}]{Cassatella+12}
Cassatella, P., Uttley, P., Wilms, J., \& Poutanen, J. 2012, MNRAS, 422, 2407

\bibitem[{Castro-Tirado {et~al.}(1992)Castro-Tirado, Brandt, \&
  Lund}]{CastroTirado+92}
Castro-Tirado, A.~J., Brandt, S., \& Lund, N. 1992, IAU Circ., 5590, 2

\bibitem[{Churazov {et~al.}(2001)Churazov, Gilfanov, \&
  Revnivtsev}]{Churazov+01}
Churazov, E., Gilfanov, M., \& Revnivtsev, M. 2001, MNRAS, 321, 759

\bibitem[{Cui {et~al.}(1997)Cui, Zhang, Jahoda, Focke, Swank, Heindl, \&
  Rothschild}]{Cui+97}
Cui, W., Zhang, S.~N., Jahoda, K., Focke, W., Swank, J.~H., Heindl, W.~A., \&
  Rothschild, R.~E. 1997, The Transparent Universe, 382, 209

\bibitem[{Davies \& Harte(1987)}]{DaviesHarte87}
Davies, R.~B., \& Harte, D.~S. 1987, Biometrika, 74, 95

\bibitem[{Dunn {et~al.}(2010)Dunn, Fender, K{\"o}rding, Belloni, \&
  Cabanac}]{Dunn+10}
Dunn, R. J.~H., Fender, R.~P., K{\"o}rding, E.~G., Belloni, T., \& Cabanac, C.
  2010, MNRAS, 403, 61

\bibitem[{Fabian {et~al.}(2009)Fabian, Zoghbi, Ross, Uttley, Gallo, Brandt,
  Blustin, Boller, Caballero-Garcia, Larsson, Miller, Miniutti, Ponti, Reis,
  Reynolds, Tanaka, \& Young}]{Fabian+09}
Fabian, A.~C. {et~al.} 2009, Nat., 459, 540

\bibitem[{Feroci {et~al.}(2012)Feroci, Stella, van~der Klis, Courvoisier,
  Hernanz, Hudec, Santangelo, Walton, Zdziarski, Barret, Belloni, Braga,
  Brandt, Budtz-J{\o}rgensen, Campana, den Herder, Huovelin, Israel, Pohl, Ray,
  Vacchi, Zane, Argan, Attin{\`a}, Bertuccio, Bozzo, Campana, Chakrabarty,
  Costa, De~Rosa, Del~Monte, Di~Cosimo, Donnarumma, Evangelista, Haas, Jonker,
  Korpela, Labanti, Malcovati, Mignani, Muleri, Rapisarda, Rashevsky, Rea,
  Rubini, Tenzer, Wilson-Hodge, Winter, Wood, Zampa, Zampa, Abramowicz, Alpar,
  Altamirano, Alvarez, Amati, Amoros, Antonelli, Artigue, Azzarello, Bachetti,
  Baldazzi, Barbera, Barbieri, Basa, Baykal, Belmont, Boirin, Bonvicini,
  Burderi, Bursa, Cabanac, Cackett, Caliandro, Casella, Chaty, Chenevez, Coe,
  Collura, Corongiu, Covino, Cusumano, D'Amico, Dall'Osso, De~Martino,
  De~Paris, Di~Persio, Di~Salvo, Done, Dov{\v c}iak, Drago, Ertan, Fabiani,
  Falanga, Fender, Ferrando, Della Monica~Ferreira, Fraser, Frontera, Fuschino,
  Galvez, Gandhi, Giommi, Godet, G{\"o}{\u{g}}{\"u}{\c s}, Goldwurm, G{\"o}tz,
  Grassi, Guttridge, Hakala, Henri, Hermsen, Horak, Hornstrup, in't Zand,
  Isern, Kalemci, Kanbach, Karas, Kataria, Kennedy, Klochkov, Klu{\'{z}}niak,
  Kokkotas, Kreykenbohm, Krolik, Kuiper, Kuvvetli, Kylafis, Lattimer,
  Lazzarotto, Leahy, Lebrun, Lin, Lund, Maccarone, Malzac, Marisaldi,
  Martindale, Mastropietro, McClintock, McHardy, Mendez, Mereghetti, Miller,
  Mineo, Morelli, Morsink, Motch, Motta, Mu{\~n}oz-Darias, Naletto, Neustroev,
  Nevalainen, Olive, Orio, Orlandini, Orleanski, Ozel, Pacciani, Paltani,
  Papadakis, Papitto, Patruno, Pellizzoni, Petr{\'a}{\v c}ek, Petri, Petrucci,
  Phlips, Picolli, Possenti, Psaltis, Rambaud, Reig, Remillard, Rodriguez,
  Romano, Romanova, Schanz, Schmid, Segreto, Shearer, Smith, Smith, Soffitta,
  Stergioulas, Stolarski, Stuchl{\'\i}k, Tiengo, Torres, T{\"o}r{\"o}k,
  Turolla, Uttley, Vaughan, Vercellone, Waters, Watts, Wawrzaszek, Webb, Wilms,
  Zampieri, Zezas, \& Ziolkowski}]{Feroci+12}
Feroci, M. {et~al.} 2012, Experimental Astronomy, 34, 415

\bibitem[{Foster {et~al.}(1996)Foster, Waltman, Tavani, Harmon, Zhang,
  Paciesas, \& Ghigo}]{Foster+96}
Foster, R.~S., Waltman, E.~B., Tavani, M., Harmon, B.~A., Zhang, S.~N.,
  Paciesas, W.~S., \& Ghigo, F.~D. 1996, Astrophysical Journal Letters v.467,
  467, L81

\bibitem[{F{\"u}rst {et~al.}(2013)F{\"u}rst, Grefenstette, Staubert, Tomsick,
  Bachetti, Barret, Bellm, Boggs, Chenevez, Christensen, Craig, Hailey,
  Harrison, Klochkov, Madsen, Pottschmidt, Stern, Walton, Wilms, \&
  Zhang}]{Fuerst+13}
F{\"u}rst, F. {et~al.} 2013, ApJ, 779, 69

\bibitem[{F{\"u}rst {et~al.}(2014)F{\"u}rst, Pottschmidt, Wilms, Tomsick,
  Bachetti, Boggs, Christensen, Craig, Grefenstette, Hailey, Harrison, Madsen,
  Miller, Stern, Walton, \& Zhang}]{Fuerst+14vela}
------. 2014, ApJ, 780, 133

\bibitem[{Gandhi {et~al.}(2010)Gandhi, Dhillon, Durant, Fabian, Kubota,
  Makishima, Malzac, Marsh, Miller, Shahbaz, Spruit, \& Casella}]{Gandhi+10}
Gandhi, P. {et~al.} 2010, MNRAS, 407, 2166

\bibitem[{Gentle(2003)}]{Gentle03}
Gentle, J.~E. 2003, {Random Number Generation and Monte Carlo Methods}
  (Springer)

\bibitem[{Gilfanov {et~al.}(2000)Gilfanov, Churazov, \&
  Revnivtsev}]{Gilfanov+00}
Gilfanov, M., Churazov, E., \& Revnivtsev, M. 2000, MNRAS, 316, 923

\bibitem[{Gleissner {et~al.}(2004)Gleissner, Wilms, Pottschmidt, Uttley, Nowak,
  \& Staubert}]{Gleissner+04}
Gleissner, T., Wilms, J., Pottschmidt, K., Uttley, P., Nowak, M.~A., \&
  Staubert, R. 2004, A{\&}A, 414, 1091

\bibitem[{Grinberg {et~al.}(2014)Grinberg, Pottschmidt, B{\"o}ck, Schmid,
  Nowak, Uttley, Tomsick, Rodriguez, Hell, Markowitz, Bodaghee, Cadolle~Bel,
  Rothschild, \& Wilms}]{Grinberg+14}
Grinberg, V. {et~al.} 2014, A{\&}A, 565, 1

\bibitem[{Harrison {et~al.}(2013)Harrison, Craig, Christensen, Hailey, Zhang,
  Boggs, Stern, Cook, Forster, Giommi, Grefenstette, Kim, Kitaguchi, Koglin,
  Madsen, Mao, Miyasaka, Mori, Perri, Pivovaroff, Puccetti, Rana, Westergaard,
  Willis, Zoglauer, An, Bachetti, Barri{\`e}re, Bellm, Bhalerao, Brejnholt,
  Fuerst, Liebe, Markwardt, Nynka, Vogel, Walton, Wik, Alexander, Cominsky,
  Hornschemeier, Hornstrup, Kaspi, Madejski, Matt, Molendi, Smith, Tomsick,
  Ajello, Ballantyne, Balokovi{\'c}, Barret, Bauer, Blandford, Brandt,
  Brenneman, Chiang, Chakrabarty, Chenevez, Comastri, Dufour, Elvis, Fabian,
  Farrah, Fryer, Gotthelf, Grindlay, Helfand, Krivonos, Meier, Miller,
  Natalucci, Ogle, Ofek, Ptak, Reynolds, Rigby, Tagliaferri, Thorsett,
  Treister, \& Urry}]{nustar13}
Harrison, F.~A. {et~al.} 2013, ApJ, 770, 103

\bibitem[{Jahoda {et~al.}(2006)Jahoda, Markwardt, Radeva, Rots, Stark, Swank,
  Strohmayer, \& Zhang}]{Jahoda+06}
Jahoda, K., Markwardt, C.~B., Radeva, Y., Rots, A.~H., Stark, M.~J., Swank,
  J.~H., Strohmayer, T.~E., \& Zhang, W. 2006, ApJ Supp. Ser., 163, 401

\bibitem[{Jin {et~al.}(2013)Jin, Done, Middleton, \& Ward}]{Jin+13}
Jin, C., Done, C., Middleton, M., \& Ward, M. 2013, MNRAS, 436, 3173

\bibitem[{Jones {et~al.}(2008)Jones, Levine, Morgan, \& Rappaport}]{Jones+08}
Jones, T.~A., Levine, A.~M., Morgan, E.~H., \& Rappaport, S. 2008, ApJ, 677,
  1241

\bibitem[{Kaspi {et~al.}(2014)Kaspi, Archibald, Bhalerao, Dufour, Gotthelf, An,
  Bachetti, Beloborodov, Boggs, Christensen, Craig, Grefenstette, Hailey,
  Harrison, Kennea, Kouveliotou, Madsen, Mori, Markwardt, Stern, Vogel, \&
  Zhang}]{Kaspi+14}
Kaspi, V.~M. {et~al.} 2014, ApJ, 786, 84

\bibitem[{K{\"o}rding \& Falcke(2004)}]{KoerdingFalcke04}
K{\"o}rding, E., \& Falcke, H. 2004, A{\&}A, 414, 795

\bibitem[{Leahy {et~al.}(1983)Leahy, Darbro, Elsner, Weisskopf, Kahn,
  Sutherland, \& Grindlay}]{Leahy+83}
Leahy, D.~A., Darbro, W., Elsner, R.~F., Weisskopf, M.~C., Kahn, S.,
  Sutherland, P.~G., \& Grindlay, J.~E. 1983, ApJ, 266, 160

\bibitem[{Lewin {et~al.}(1988)Lewin, van Paradijs, \& van~der Klis}]{Lewin+88}
Lewin, W. H.~G., van Paradijs, J., \& van~der Klis, M. 1988, SSRv, 46, 273

\bibitem[{Maccarone \& Coppi(2003)}]{MaccaroneCoppi03}
Maccarone, T.~J., \& Coppi, P.~S. 2003, MNRAS, 338, 189

\bibitem[{Marinucci {et~al.}(2014)Marinucci, Matt, Kara, Miniutti, Elvis,
  Ar{\'e}valo, Ballantyne, Balokovi{\'c}, Bauer, Brenneman, Boggs, Cappi,
  Christensen, Craig, Fabian, Fuerst, Hailey, Harrison, Risaliti, Reynolds,
  Stern, Walton, \& Zhang}]{Marinucci+14}
Marinucci, A. {et~al.} 2014, MNRAS, 440, 2347

\bibitem[{Markert {et~al.}(1973)Markert, Canizares, Clark, Lewin, Schnopper, \&
  Sprott}]{Markert+73}
Markert, T.~H., Canizares, C.~R., Clark, G.~W., Lewin, W. H.~G., Schnopper,
  H.~W., \& Sprott, G.~F. 1973, ApJ, 184, L67

\bibitem[{McClintock \& Remillard(2006)}]{McClintockRemillard06}
McClintock, J.~E., \& Remillard, R.~A. 2006, in Compact stellar X-ray sources,
  ed. J.~E. McClintock \& R.~A. Remillard, 157--213

\bibitem[{Miller {et~al.}(2013)Miller, Parker, Fuerst, Bachetti, Harrison,
  Barret, Boggs, Chakrabarty, Christensen, Craig, Fabian, Grefenstette, Hailey,
  King, Stern, Tomsick, Walton, \& Zhang}]{Miller+131915}
Miller, J.~M. {et~al.} 2013, ApJL, 775, L45

\bibitem[{Miyamoto {et~al.}(1991)Miyamoto, Kimura, Kitamoto, Dotani, \&
  Ebisawa}]{Miyamoto+91}
Miyamoto, S., Kimura, K., Kitamoto, S., Dotani, T., \& Ebisawa, K. 1991, ApJ,
  383, 784

\bibitem[{Miyamoto \& Kitamoto(1989)}]{Miyamoto+89}
Miyamoto, S., \& Kitamoto, S. 1989, Nat., 342, 773

\bibitem[{Miyasaka {et~al.}(2013)Miyasaka, Bachetti, Harrison, F{\"u}rst,
  Barret, Bellm, Boggs, Chakrabarty, Chenevez, Christensen, Craig,
  Grefenstette, Hailey, Madsen, Natalucci, Pottschmidt, Stern, Tomsick, Walton,
  Wilms, \& Zhang}]{Miyasaka+13}
Miyasaka, H. {et~al.} 2013, ApJ, 775, 65

\bibitem[{Morgan {et~al.}(1997)Morgan, Remillard, \& Greiner}]{Morgan+97}
Morgan, E.~H., Remillard, R.~A., \& Greiner, J. 1997, ApJ, 482, 993

\bibitem[{Mori {et~al.}(2013)Mori, Gotthelf, Zhang, An, Baganoff, Barri{\`e}re,
  Beloborodov, Boggs, Christensen, Craig, Dufour, Grefenstette, Hailey,
  Harrison, Hong, Kaspi, Kennea, Madsen, Markwardt, Nynka, Stern, Tomsick, \&
  Zhang}]{Mori+13}
Mori, K. {et~al.} 2013, ApJL, 770, L23

\bibitem[{Natalucci {et~al.}(2014)Natalucci, Tomsick, Bazzano, Smith, Bachetti,
  Barret, Boggs, Christensen, Craig, Fiocchi, F{\"u}rst, Grefenstette, Hailey,
  Harrison, Krivonos, Kuulkers, Miller, Pottschmidt, Stern, Ubertini, Walton,
  \& Zhang}]{Natalucci+14}
Natalucci, L. {et~al.} 2014, ApJ, 780, 63

\bibitem[{Nowak \& Vaughan(1996)}]{Nowak+96}
Nowak, M.~A., \& Vaughan, B.~A. 1996, MNRAS, 280, 227

\bibitem[{Nowak {et~al.}(1999{\natexlab{a}})Nowak, Vaughan, Wilms, Dove, \&
  Begelman}]{Nowak+99}
Nowak, M.~A., Vaughan, B.~A., Wilms, J., Dove, J.~B., \& Begelman, M.~C.
  1999{\natexlab{a}}, ApJ, 510, 874

\bibitem[{Nowak {et~al.}(1999{\natexlab{b}})Nowak, Wilms, \&
  Dove}]{Nowak+99339}
Nowak, M.~A., Wilms, J., \& Dove, J.~B. 1999{\natexlab{b}}, ApJ, 517, 355

\bibitem[{Pahari {et~al.}(2013)Pahari, Neilsen, Yadav, Misra, \&
  Uttley}]{Pahari+13}
Pahari, M., Neilsen, J., Yadav, J.~S., Misra, R., \& Uttley, P. 2013, ApJ, 778,
  136

\bibitem[{Papadakis {et~al.}(2001)Papadakis, Nandra, \& Kazanas}]{Papadakis+01}
Papadakis, I.~E., Nandra, K., \& Kazanas, D. 2001, ApJ, 554, L133

\bibitem[{Paul \& Team(2009)}]{Paul09}
Paul, B., \& Team, L. 2009, in Astrophysics with All-Sky X-Ray Observations,
  362

\bibitem[{Pottschmidt {et~al.}(2000)Pottschmidt, Wilms, Nowak, Heindl, Smith,
  \& Staubert}]{Pottschmidt+00}
Pottschmidt, K., Wilms, J., Nowak, M.~A., Heindl, W.~A., Smith, D.~M., \&
  Staubert, R. 2000, A{\&}A, 357, L17

\bibitem[{Reig {et~al.}(2000)Reig, Belloni, van~der Klis, Mendez, Kylafis, \&
  Ford}]{Reig+00}
Reig, P., Belloni, T., van~der Klis, M., Mendez, M., Kylafis, N.~D., \& Ford,
  E.~C. 2000, ApJ, 541, 883

\bibitem[{Remillard {et~al.}(2003)Remillard, Muno, McClintock, \&
  Orosz}]{Remillard+03}
Remillard, R.~A., Muno, M.~P., McClintock, J.~E., \& Orosz, J.~A. 2003, AAS, 7,
  648

\bibitem[{Revnivtsev {et~al.}(1999)Revnivtsev, Gilfanov, \&
  Churazov}]{Revnivtsev+99}
Revnivtsev, M., Gilfanov, M., \& Churazov, E. 1999, A{\&}A, 347, L23

\bibitem[{Revnivtsev {et~al.}(2000)Revnivtsev, Gilfanov, \&
  Churazov}]{Revnivtsev+00}
------. 2000, A{\&}A, 363, 1013

\bibitem[{Risaliti {et~al.}(2013)Risaliti, Harrison, Madsen, Walton, Boggs,
  Christensen, Craig, Grefenstette, Hailey, Nardini, Stern, \&
  Zhang}]{Risaliti+13}
Risaliti, G. {et~al.} 2013, Nat., 494, 449

\bibitem[{Strohmayer(2001)}]{Strohmayer01}
Strohmayer, T.~E. 2001, ApJ, 554, L169

\bibitem[{Timmer \& Koenig(1995)}]{TimmerKoenig95}
Timmer, J., \& Koenig, M. 1995, A{\&}A, 300, 707

\bibitem[{Tomsick \& Kaaret(2001)}]{TomsickKaaret01}
Tomsick, J.~A., \& Kaaret, P. 2001, ApJ, 548, 401

\bibitem[{Tomsick {et~al.}(2014)Tomsick, Nowak, Parker, Miller, Fabian,
  Harrison, Bachetti, Barret, Boggs, Christensen, Craig, Forster, F{\"u}rst,
  Grefenstette, Hailey, King, Madsen, Natalucci, Pottschmidt, Ross, Stern,
  Walton, Wilms, \& Zhang}]{Tomsick+14}
Tomsick, J.~A. {et~al.} 2014, ApJ, 780, 78

\bibitem[{Uttley {et~al.}(2014)Uttley, Cackett, Fabian, Kara, \&
  Wilkins}]{Uttley+14}
Uttley, P., Cackett, E.~M., Fabian, A.~C., Kara, E., \& Wilkins, D.~R. 2014,
  arXiv, 6575, 1405.6575

\bibitem[{Uttley {et~al.}(2011)Uttley, Wilkinson, Cassatella, Wilms,
  Pottschmidt, Hanke, \& B{\"o}ck}]{Uttley+11}
Uttley, P., Wilkinson, T., Cassatella, P., Wilms, J., Pottschmidt, K., Hanke,
  M., \& B{\"o}ck, M. 2011, MNRAS Let., 414, L60

\bibitem[{van~der Klis(1989)}]{VDK89}
van~der Klis, M. 1989, in Timing Neutron Stars: proceedings of the NATO
  Advanced Study Institute on Timing Neutron Stars held April 4-15, 27

\bibitem[{Vaughan \& Nowak(1997)}]{Vaughan+97}
Vaughan, B.~A., \& Nowak, M.~A. 1997, ApJL, 474, L43

\bibitem[{Vaughan(2013)}]{Vaughan13}
Vaughan, S. 2013, arXiv, 1309.6435v1

\bibitem[{Vikhlinin {et~al.}(1994)Vikhlinin, Churazov, \&
  Gilfanov}]{Vikhlinin+94}
Vikhlinin, A., Churazov, E., \& Gilfanov, M. 1994, A{\&}A, 287, 73

\bibitem[{Walton {et~al.}(2014)Walton, Risaliti, Harrison, Fabian, Miller,
  Ar{\'e}valo, Ballantyne, Boggs, Brenneman, Christensen, Craig, Elvis, Fuerst,
  Gandhi, Grefenstette, Hailey, Kara, Luo, Madsen, Marinucci, Matt, Parker,
  Reynolds, Rivers, Ross, Stern, \& Zhang}]{Walton+141365}
Walton, D.~J. {et~al.} 2014, arXiv, 5620, 1404.5620

\bibitem[{Weisskopf(1985)}]{Weisskopf85}
Weisskopf, M.~C. 1985, in {Talk presented at the Workshop {\it Time Variability
  in X-ray and Gamma-Ray sources}, Taos, NM, USA}

\bibitem[{Zhang {et~al.}(1995)Zhang, Jahoda, Swank, Morgan, \&
  Giles}]{Zhang+95}
Zhang, W., Jahoda, K., Swank, J.~H., Morgan, E.~H., \& Giles, A.~B. 1995, ApJ,
  449, 930

\bibitem[{Zoghbi {et~al.}(2014)Zoghbi, Cackett, Reynolds, Kara, Harrison,
  Fabian, Lohfink, Matt, Balokovi{\'c}, Boggs, Christensen, Craig, Hailey,
  Stern, \& Zhang}]{Zoghbi+14}
Zoghbi, A. {et~al.} 2014, ApJ, 789, 56

\end{thebibliography}
\end{document}